\documentclass[prl,aps,twocolumn,superscriptaddress,longbibliography]{revtex4-1}
\usepackage{nopageno}
\usepackage{latexsym}
\usepackage{lineno}
\usepackage{hyperref}
\usepackage{url}
\usepackage{graphicx}
\usepackage{verbatim}
\usepackage{multirow}
\usepackage{amsmath, bm}
\newcommand{\beq}{\begin{eqnarray}}
\newcommand{\eeq}{\end{eqnarray}}
\usepackage{mathrsfs}
\usepackage{float}
\usepackage[usenames, dvipsnames]{color}
\usepackage{mathtools}
\usepackage{slashed}
\usepackage{physics}	% installed packages for typing maths and physics
\usepackage{graphicx}   % need for Fig.ures
\usepackage{epstopdf}
\usepackage{subfigure}  % use for side-by-side Fig.ures
\usepackage{hyperref}   % use for hypertext links, including those to external documents and URLs
\usepackage{bbold}
\usepackage{wasysym}
\usepackage{amssymb}
\usepackage{feynmp}
\usepackage[symbol]{footmisc}
\def \bs{\textbf}
\usepackage[latin1]{inputenc}
\usepackage{tikz}
\usetikzlibrary{decorations.pathmorphing}
\usetikzlibrary{shapes.misc}
\tikzset{cross/.style={cross out, draw=black, minimum size=8*(#1-\pgflinewidth), inner sep=0pt, outer sep=0pt},
cross/.default={1pt}}
\usetikzlibrary{patterns,math}
\newcommand{\RN}[1]{%
  \textup{\uppercase\expandafter{\romannumeral#1}}%
}

\newcommand{\qs}[1]{{\color{black} #1}}

\newcommand{\setty}[1]{{\color{black} #1}}
\begin{document}

\title{Electron correlations and charge density wave in the topological kagome metal FeGe
}
\author{Chandan Setty$^{\oplus,\dagger}$}
 \affiliation{%
Department of Physics \& Astronomy,
Rice Center for Quantum Materials,
Rice University, Houston, Texas 77005,USA
 }
\author{Christopher A. Lane $^{\oplus}$}
\affiliation{Theoretical Division, Los Alamos National Laboratory, Los Alamos, New Mexico 87545 USA}
\author{Lei Chen}
 \affiliation{%
Department of Physics \& Astronomy,
Rice Center for Quantum Materials,
Rice University, Houston, Texas 77005,USA
 }
\author{Haoyu Hu}
 \affiliation{%
Department of Physics \& Astronomy,
Rice Center for Quantum Materials,
Rice University, Houston, Texas 77005,USA
 }
\author{Jian-Xin Zhu
}
\affiliation{Theoretical Division, Los Alamos National Laboratory, Los Alamos, New Mexico 87545 USA}
 \author{Qimiao Si
 }
 \affiliation{%
Department of Physics \& Astronomy,
Rice Center for Quantum Materials,
Rice University, Houston, Texas 77005,USA
 }%

\begin{abstract}
Charge order in kagome metals is  of extensive current interest. Recently, a charge
density wave was discovered in the \textit{magnetic} binary kagome metal FeGe~\cite{Dai2022}.
In analogy to its predecessor, the non-magnetic 
$A$V$_3$Sb$_5$ ($A$=K, Cs, Rb),
the in-plane ordering occurs at the $M$ point. 
 In contrast, however, the system 
 manifestly 
 shows effects of substantial 
 correlations.
  Here we identify the topological bands  
 crossing the Fermi energy (E$_F$)
  in FeGe
  and 
  characterize the correlation-induced renormalization of these bands.
We then
 derive
  a charge order from 
an effective model comprising 
 topological kagome `flat' bands
in the presence of a magnetic order.
We
 demonstrate 
 edge states as well as 
 excess
out-of-plane magnetic moment 
\qs{associated with}
 the charge 
order; both are fingerprints of non-trivial band topology and 
consistent with recent
experimental observations.
Our results point to FeGe as an ideal platform to 
realize and elucidate
\qs{correlated topological physics.}

\end{abstract}
\maketitle
\textit{Introduction:} Charge order phases in kagome metals have garnered much 
recent
attention
owing to their highly unconventional properties. 
Especially in
the vanadium based non-magnetic ternary kagome materials ($A$V$_3$Sb$_5$; $A$=K, Cs, Rb) ~\cite{Ortiz2019, Ortiz2020, Ortiz2021}, 
charge order has been 
found~\cite{Hasan2020, Zeljkovic2021, Wen2021, XHChen2021, Tsirlin2021, Miao2021, Shi2021}
\qs{and} shows
 several non-trivial characteristics.
 \qs{These include the breaking of time-reversal (T) symmetry}
 ~\cite{Guguchia2021-TRSB, Zhao2021-TRSB, Wang2021-Kerr}, 
  anomalous Hall transport~\cite{Ali2020-2}, 
  nematicity~\cite{Zeljkovic2021-Nematic, Wang2021-Kerr} and 
  other accompanying/competing orders~\cite{Wilson2021-doping, XHChen2021, Miao2021, Blumberg2022},  
  and possible coexistence with unconventional superconductivity~\cite{Gao2021}. \par
Another 
prominent class of transition element kagome metals
is the binary T$_m$X$_n$ (T= Fe, Co, Ni and X= Ge, Sn).
They have magnetic
ground states~\cite{Checkelsky2018, Wang2020, Zhang2018, Felser2018, Yao18.1x} and
possess  flat bands near the Fermi level (E$_F$)~\cite{Comin2020, Comin2020-2, Wang2020}.   
Recently hexagonal FeGe (SG 191, P6/$mmm$) was shown to be the first 
member in this family ($T_m \sim$ 400 K~\cite{Kunitomi1966}) 
where charge order 
is identified, with  a transition
temperature 
of $T_{CO}\sim 110$ K~\cite{Dai2022,Yin2022,Yi2022}.
The 
spins are aligned (anti-)ferromagnetically (between) within the kagome planes and 
point in a direction perpendicular to the layers ~\cite{Kunitomi1966}.

Much of the currently known phenomenology of the charge order is analogous to the ternary compounds 
-- the in-plane ordering occurs at $M$ point ($2\times2$ order) 
and the STM peaks are sensitive to the directionality of the applied perpendicular ($c$-axis) magnetic field.
However,
in contrast with the ternary family,
here the charge order emerges from a magnetic parent 
phase~\cite{Dai2022},
 and is accompanied by edge states~\cite{Yin2022} and
 \qs{an excess contribution to the ordered}
 magnetic moment~\cite{Dai2022} below $T_{CO}$. \par
In
 this work we develop a 
 model of charge order in FeGe. We identify two minimal ingredients -- the presence of flat bands crossing E$_F$ and their non-trivial topology -- that can broadly capture much of the observed data. We justify the relevance of these two ingredients to FeGe using first principle calculations. 
 To isolate the salient physics from the complexities of the real-material 
 \qs{electronic}
  structure,
 we begin from an extended Hubbard model on a kagome lattice. 
 We then seek a low energy model for the topological flat bands by working in the limit where Coulomb interactions 
 are comparable to the bandwidth of the flat bands. By solving the low-energy model non-perturbatively 
 using the slave-spin method, we analyze the nature of the resulting charge order phase with focus on the role played 
 by the parent magnetic phase.  \textcolor{black}{We find that the non-trivial band topology of the kagome lattice,  
 and its interplay with charge order and magnetism give rise to topological edge states in the charge order gap.  
\qs{Another}
key result we uncover is an
 excess
  magnetic moment, 
  which points along the existing 
magnetic
order,  arising from the intertwining of charge and spin degrees of freedom -- a property unique to correlated phases derived from topological bands~\cite{Setty2021}. Thus we identify previously unknown  fingerprints of non-trivial band topology manifest in correlated phases. 
}
\par 
Our strategy is to isolate
a pair of flat bands crossing E$_F$ with chern numbers $\pm 1$ that are split by the 
internal
 field of the parent magnetic phase. To write an effective model with interactions for this pair of `active' bands, we follow the framework
  developed by 
  several of
  us in Ref.~\cite{Setty2021} where such a charge order was already 
  predicted 
  for binary kagome metals.
 Wannier construction of exponentially localized wannier orbitals (WOs)~\cite{Vanderbilt2011, Vanderbilt2012, Vanderbilt2012-RMP, Kunevs2010} allows for tight binding description of the low energy topological bands,
from which we can study the correlation effect.
 \par 
\begin{figure}[t!]
\includegraphics[width=1.65in,height=1.6in]{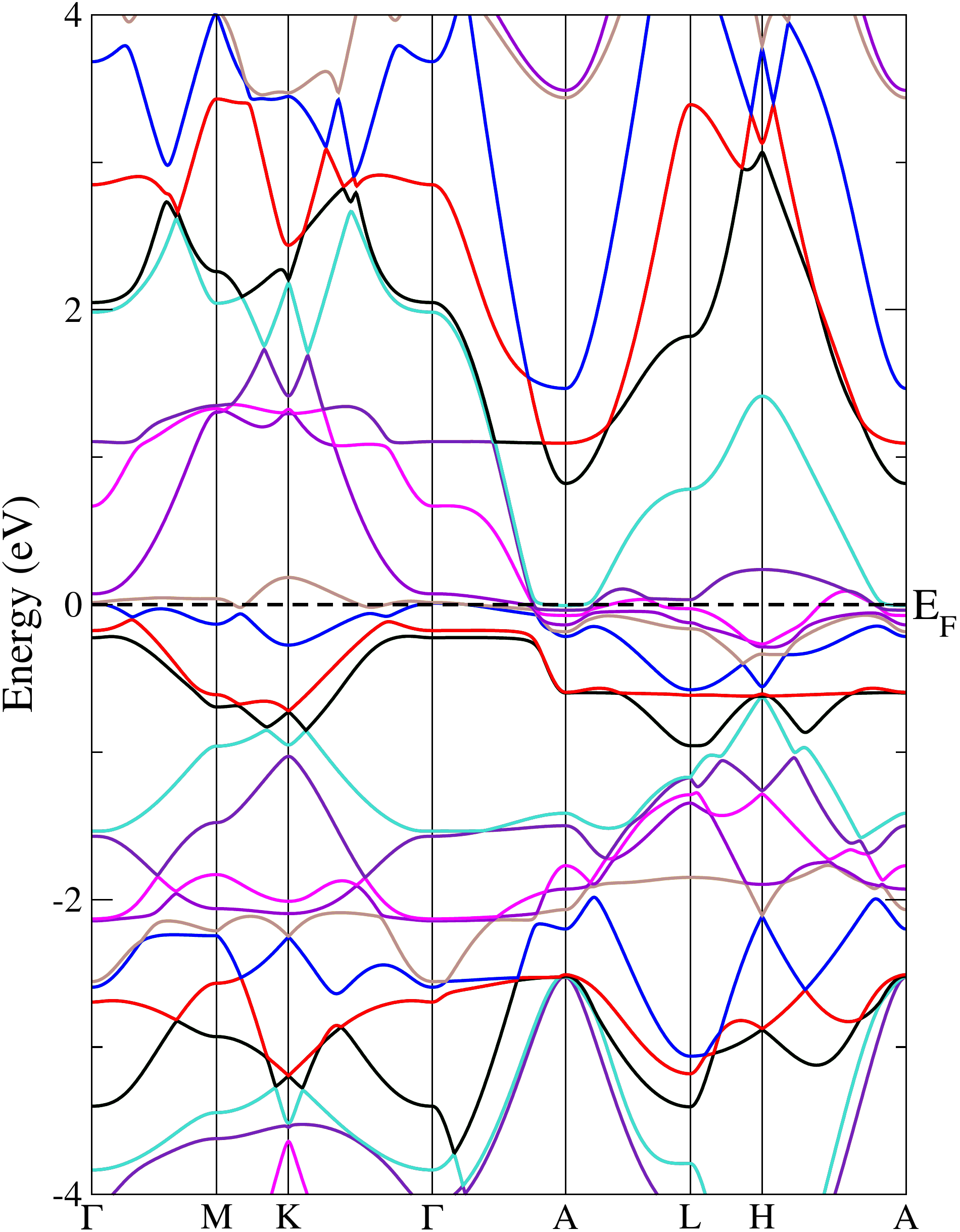} 
\includegraphics[width=1.65in,height=1.6in]{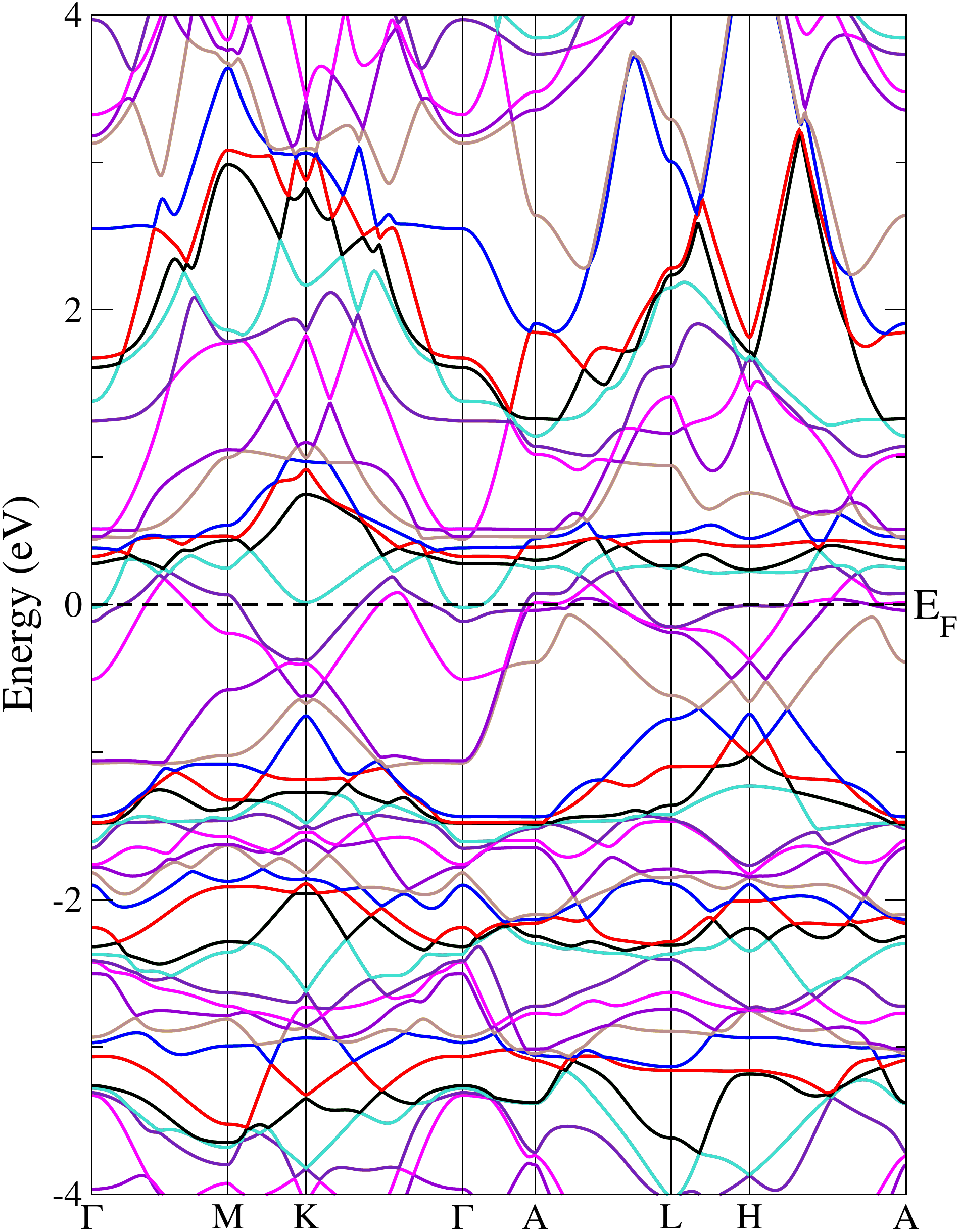}
\includegraphics[width=1.65in,height=1.6in]{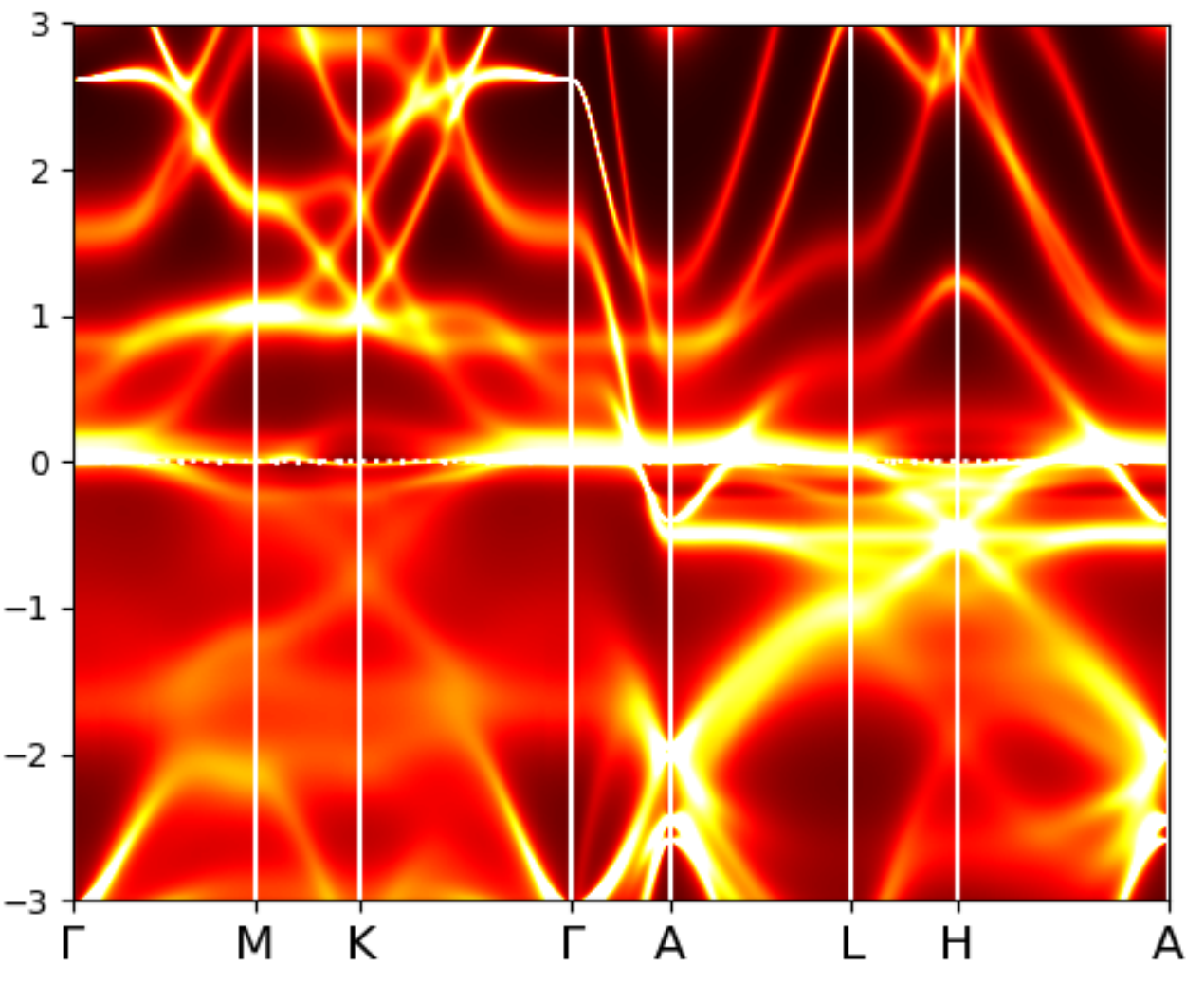} 
\includegraphics[width=1.65in,height=1.6in]{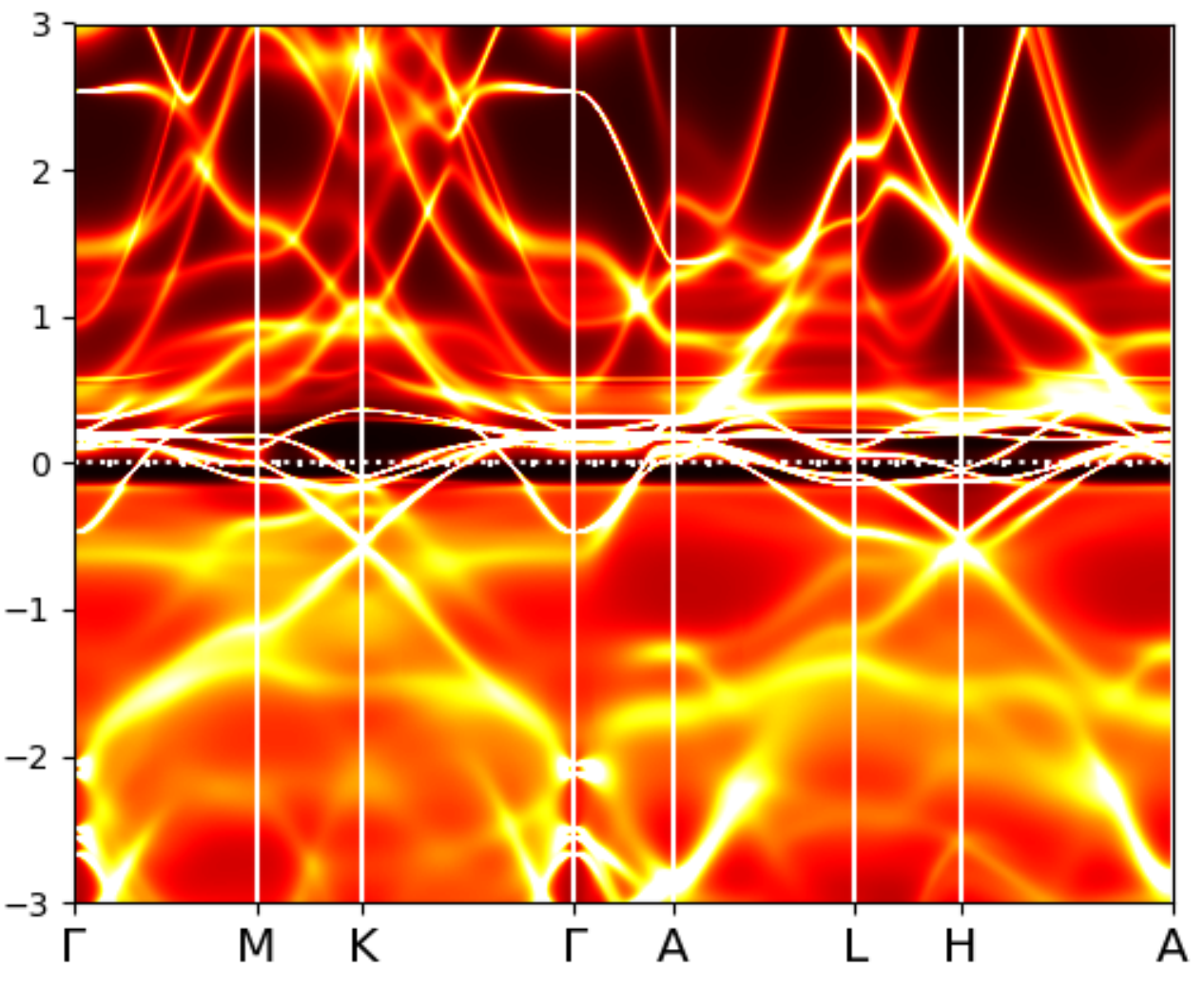}
\caption{ (Top row) First principles band structure of FeGe in the nonmagnetic (left) and magnetic (right) phases with SOC.  (Bottom row) Same as top row but with 
\qs{Coulomb}
 interactions (U = 4 eV) accounted for through DMFT.
}\label{ElectronicStructure}
%\vskip -0.3 cm
\end{figure}
\textit{Topological bands:}
Before describing the minimal interacting model, we discuss the electronic structure of FeGe and its corresponding topological characteristics.  Fig.~\ref{ElectronicStructure} (top row) shows the spin-orbit coupled
bands,
\qs{calculated by density-functional theory (DFT),}
 in the nonmagnetic 
 \qs{(NM)}
(left panel)
and magnetic
(right panel)
 phases respectively.  In
 both the nonmagnetic phase (SG P6/$mmm$) 
 and the magnetic one (SG P6/$m'm'm'$), 
 the flat bands directly cross E$_F$.
  \par
\begin{figure}[t!]
\includegraphics[width=1.65in,height=1.5in]{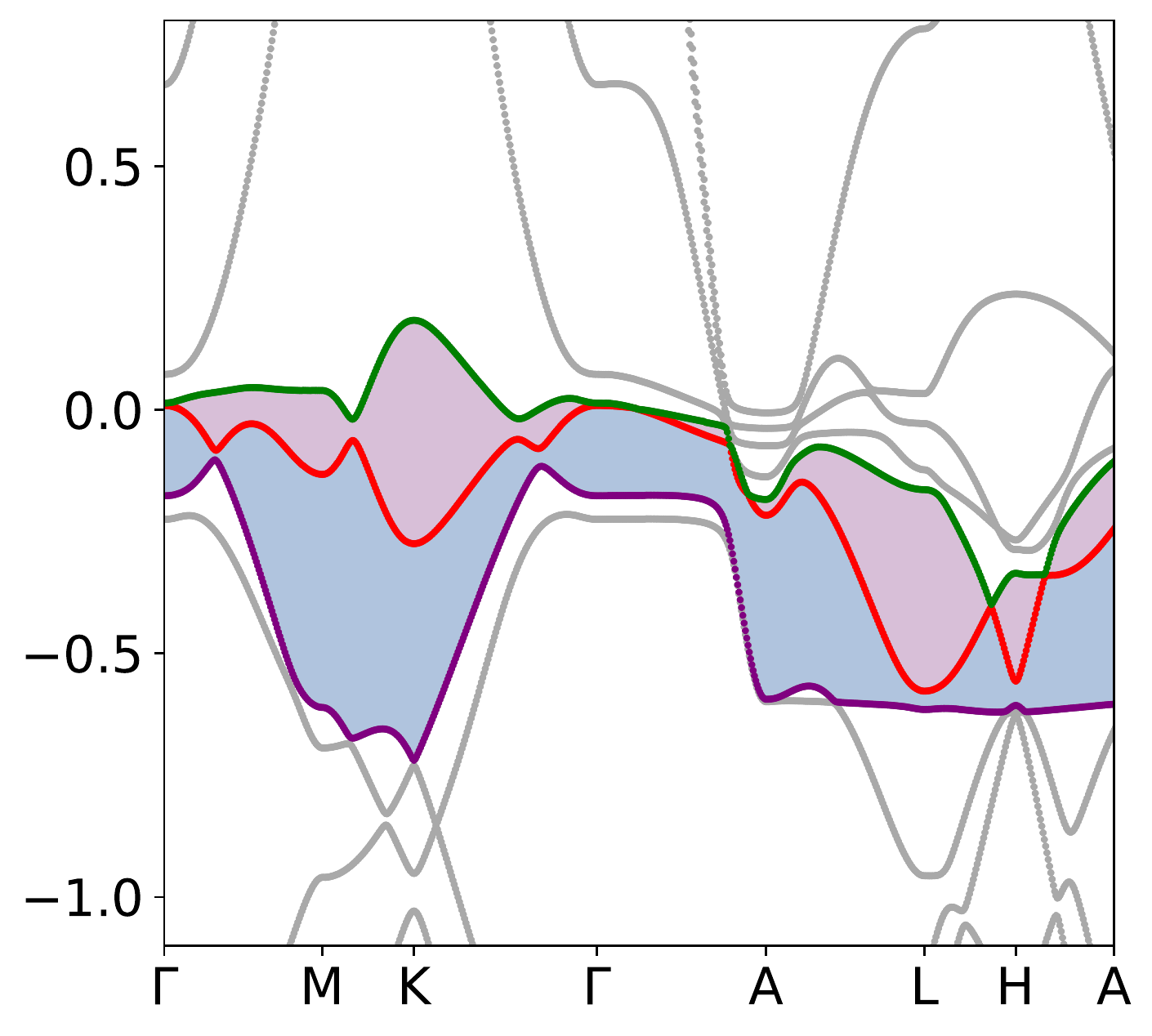} 
\includegraphics[width=1.65in,height=1.3in]{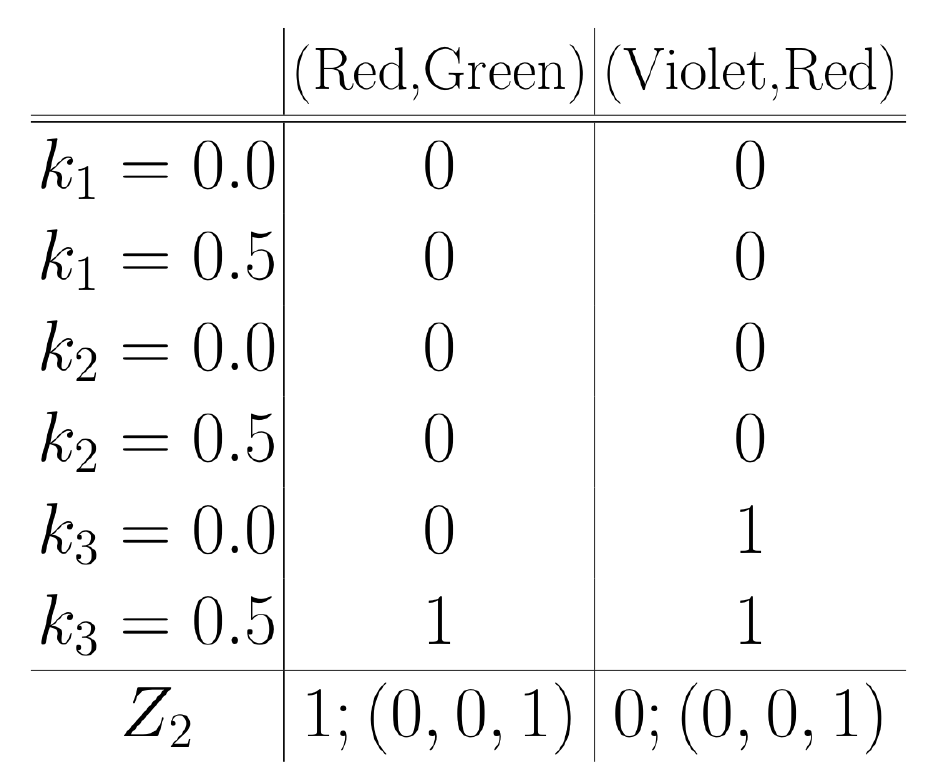}
\includegraphics[width=1.65in,height=1.4in]{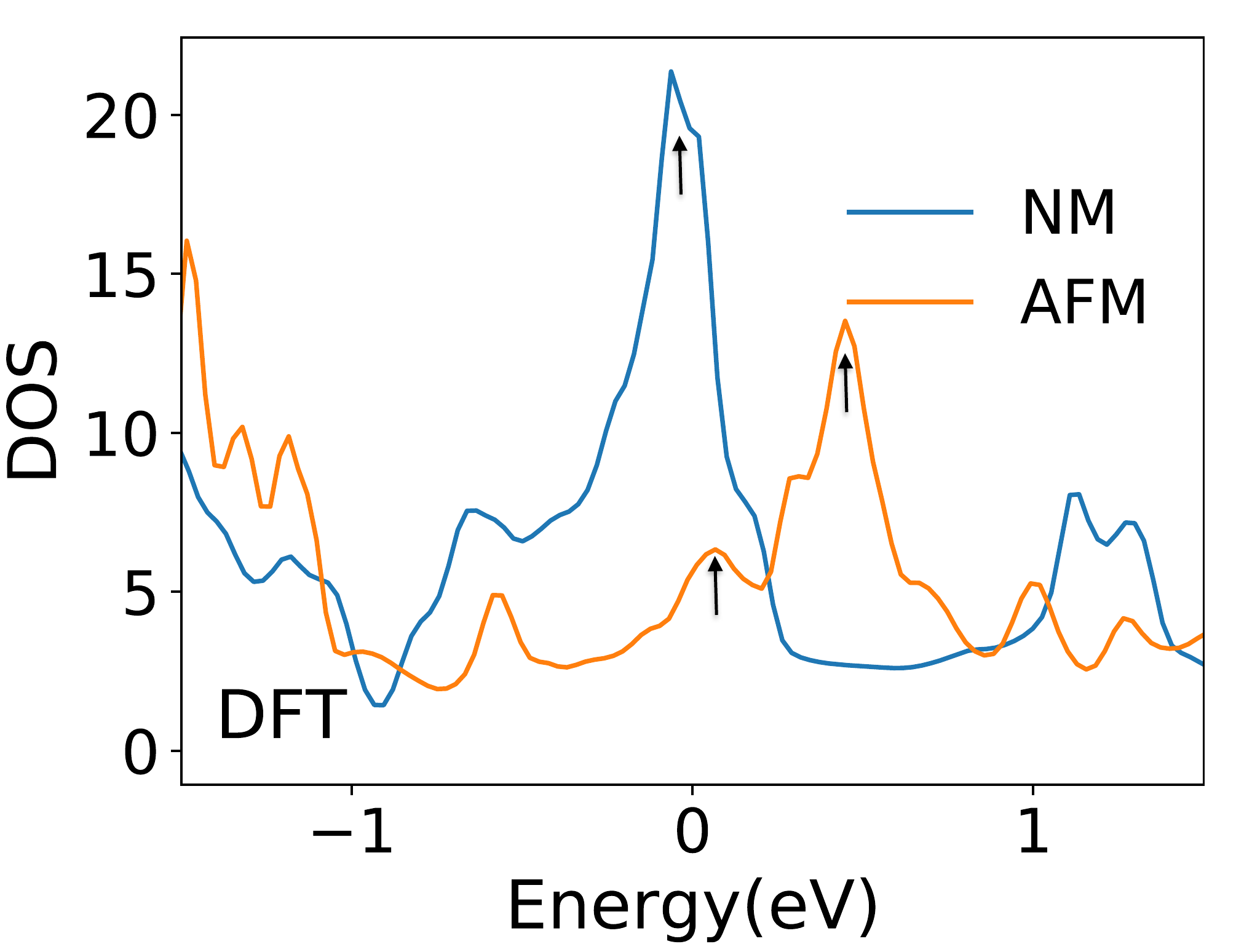}
\includegraphics[width=1.65in,height=1.4in]{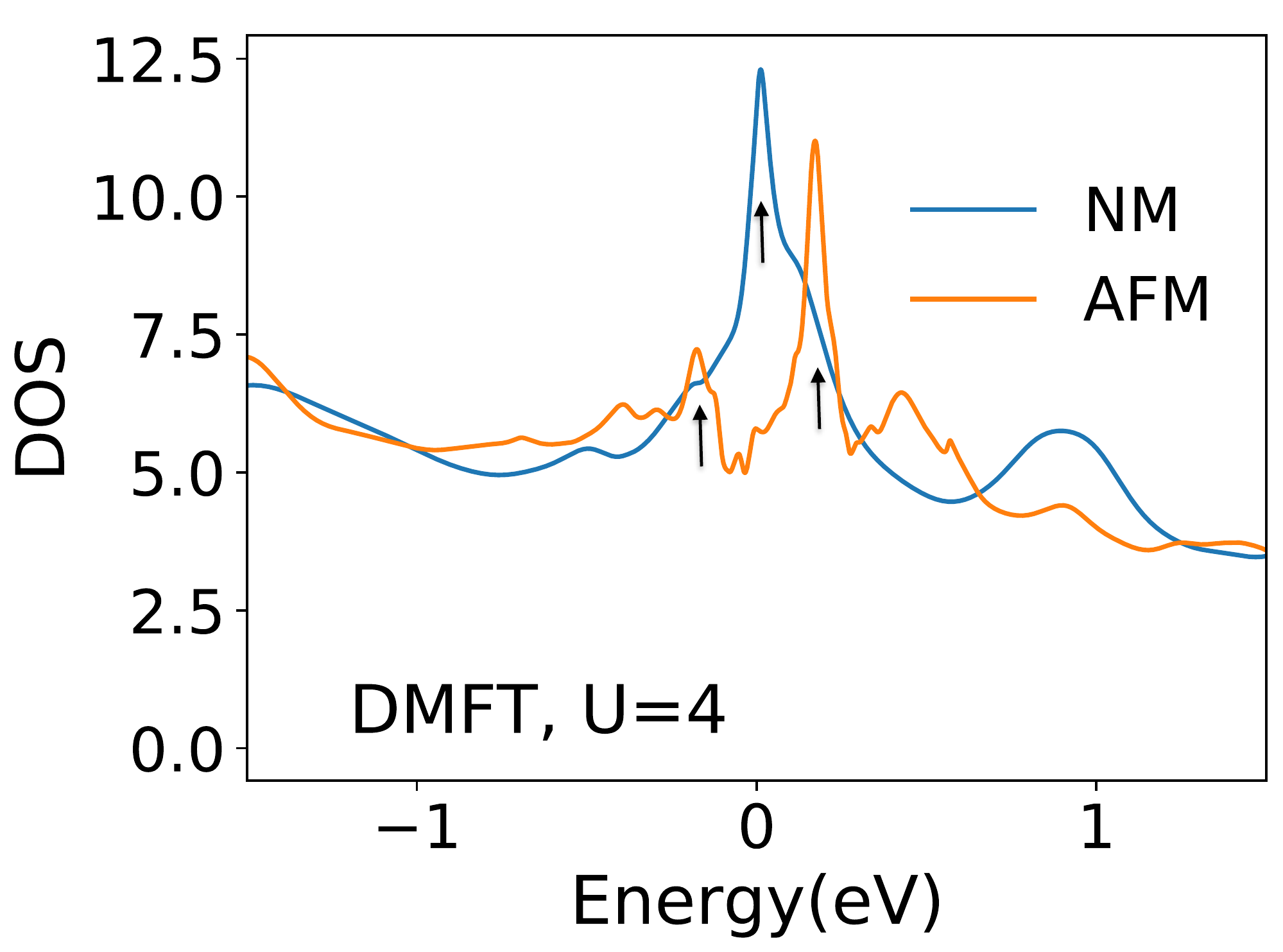} 
\caption{ \textcolor{black}{(Top left) The three topological bands near E$_F$ in the NM phase where the gaps are shaded in blue and red, (top right) $Z_2$ number for each of the six time reversal invariant planes and resulting $Z_2$ topological index for the (green, red) and (violet,red) pair of bands as shown 
\qs{in}
the top left panel. (Bottom left) DOS in the NM and magnetic phases without interactions and (Bottom right) DOS results from DMFT. The black arrows indicate flat band contributions to DOS.}
}\label{Topology}
%\vskip -0.3 cm
\end{figure}

 Fig.~\ref{Topology} (top left) highlights three topological bands near E$_F$ in the  
 \qs{NM}
 phase with spin-orbit coupling (SOC) in green, red and purple. 
 The topological indices for these
 \qs{band pairs}
  appear in Fig.~\ref{Topology} (top right). \par
 \textit{Electron correlations and DMFT calculations:}
 FeGe shows the effect of substantial electron correlations~\cite{Dai2022}. It has 
 a relatively large room temperature resistivity,
 $\sim 160 \mu \Omega\cdot$cm, 
 \qs{which reaches}
 the Mott-Ioffe-Regel value, implicating 
 FeGe as a bad metal. The electronic bands also show substantial correlation-induced renormalization~\cite{Dai2022, Yi2022}.
 We characterize this correlation effect using dynamical mean field theory (DMFT). The DFT$+$DMFT~\cite{Haule2010, Blaha2020} results are shown in
the bottom row in Fig.~\ref{ElectronicStructure} .
The flat bands become very narrow  and the spectral weight is highly concentrated at E$_F$ in both the 
\qs{NM}
and magnetic phases. See also Supplemental Material (SM) Fig.~\ref{DOS-Supp}.
\par
Importantly, the topologically non-trivial flat bands still cross E$_F$.
The 
density of states (DOS)
 near
E$_F$ 
are completely dominated by the flat bands in both the NM and magnetic phases. 
This is highlighted by black arrows in Fig.~\ref{Topology} (bottom left). 
The conclusion stands 
 in the presence of interactions as seen in DMFT (bottom right panel of Fig.~\ref{Topology}).     \par 
\begin{figure}[t!]
\includegraphics[width=1.5in,height=1.5in]{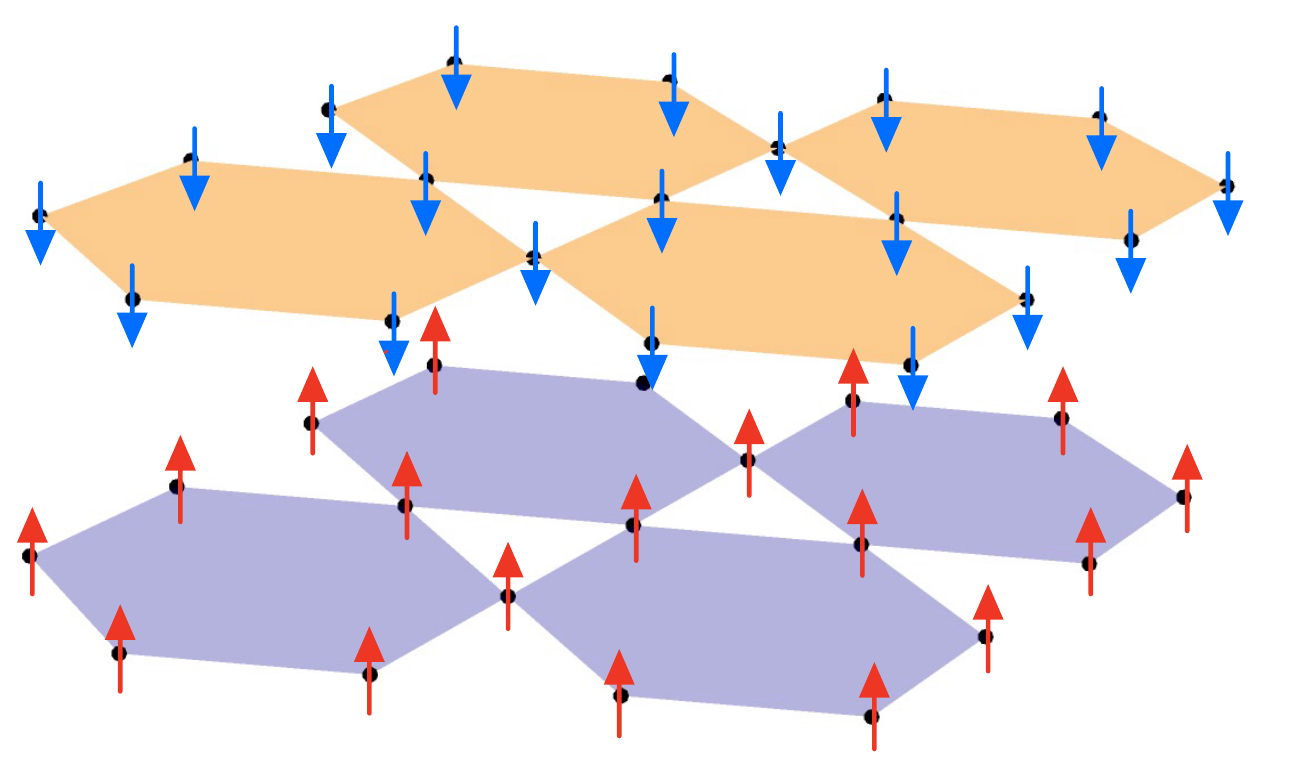}
\includegraphics[width=1.5in,height=1.5in]{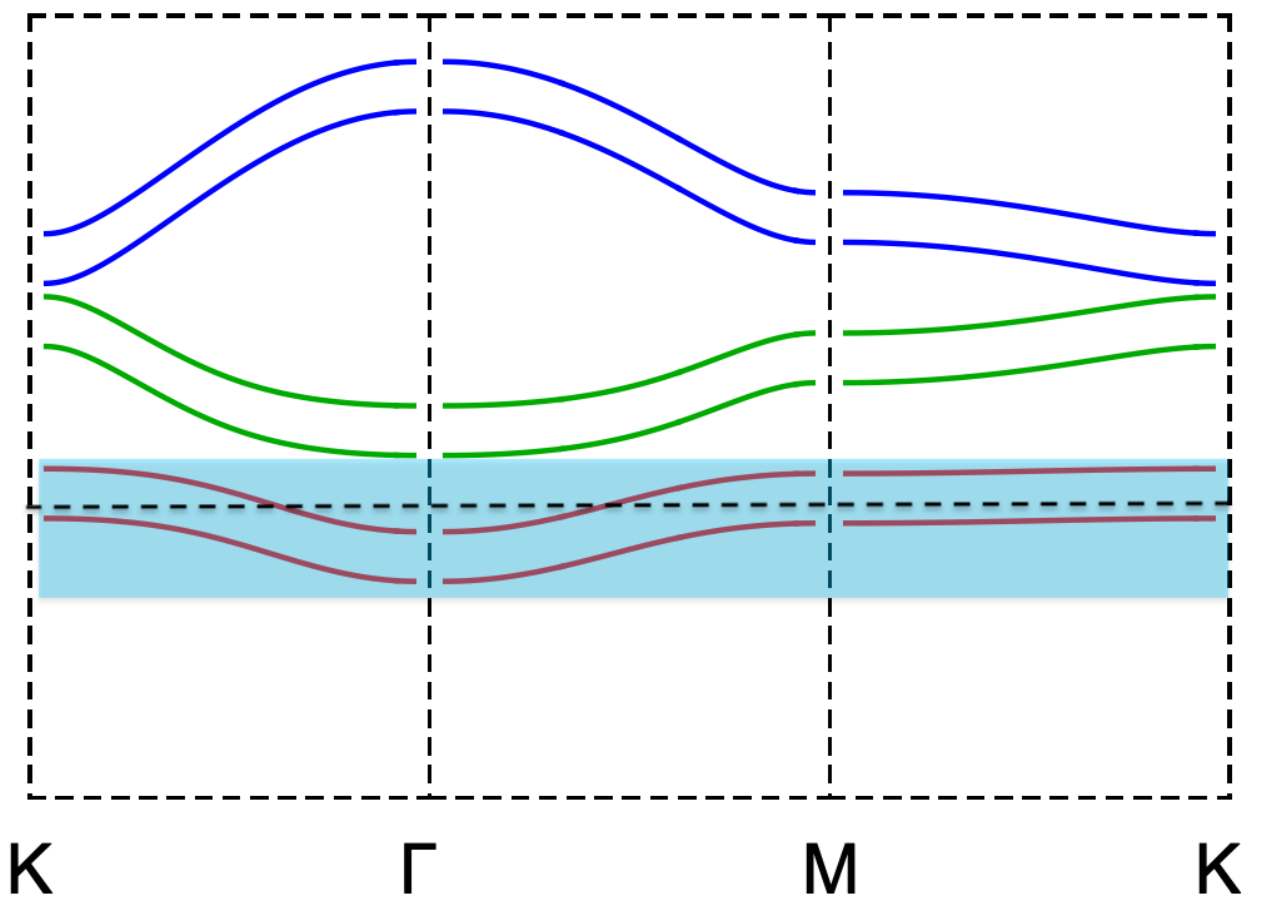}
\caption{ (Left) The kagome lattice of FeGe in the magnetic phase with ferromagnetic in-plane and anti-ferromagnetic out-of-plane moments. (Right) Band structure in the presence of spin-orbit coupling with the 'active' topological bands near E$_F$ (dashed line) marked in blue. 
}\label{kagomebands}
%\vskip -0.3 cm
\end{figure}

 \textit{Effective model for correlated topological bands:}
\setty{Having provided evidence for the existence of topological flat bands crossing E$_F$, we now utilize the minimal ingredients to study correlation-driven charge order. 
As a proof-of-principle demonstration, we focus on the 
 intralayer couplings within a kagome layer.}
The  minimal total Hamiltonian in the physical electron basis (UV limit) is given as $\mathscr{H} = \mathscr{H}_0 + \mathscr{H}_I$. The kinetic part $\mathscr{H}_0$ is described by electrons hopping on a kagome lattice in the presence of a Zeeman field of the 
magnetic
 order within the plane and is written as
\beq \nonumber
\mathscr{H}_0 &=& -t \sum_{\langle i j \rangle \alpha \beta } c_{i \alpha }^{\dagger} c_{j \beta } +  
i \lambda_1\sum_{\langle i j\rangle \alpha \beta } \left( \bs E_{ij} \times \bs R_{ij}\right) \cdot c_{i \alpha }^{\dagger} \sigma  c_{j \beta } \\ \nonumber
&&
-t_2 \sum_{\langle \langle i j \rangle \rangle \alpha \beta } c_{i \alpha }^{\dagger} c_{j \beta } +  
i \lambda_2\sum_{\langle \langle i j \rangle \rangle \alpha \beta } \left( \bs E_{ij} \times \bs R_{ij}\right) \cdot c_{i \alpha }^{\dagger} \sigma  c_{j \beta }  \\
&&+ h \sum_{i\alpha} c_{i\alpha}^{\dagger} \sigma_z c_{i\alpha}. 
\label{KagomeSOC}
\eeq
Here $c_{j\alpha}^{\dagger}$ creates an electron at site $j$ and
internal quantum number
 $\alpha$. The internal quantum numbers include
  both sub-lattice and spin indices. $t, t_2$ and $\lambda_{1,2}$ are the hopping and SOC parameters for the nearest neighbor and next-nearest neighbor lattice sites respectively.  The displacement from site $i$ to site $j$ is $\bs R_{ij}$, and the electric field experienced by the electrons along $\bs R_{ij}$ is $\bs E_{ij}$. 
  The last term arises due to the Zeeman field experienced by the electrons from the magnetic moments pointed out-of-plane in a quasi 2D kagome plane.
  We assume this field to be uniform and proportional to $h$.  Fig.~\ref{kagomebands} (left panel) describes the lattice structure and magnetically ordered moments. 
The interaction part, $\mathscr{H}_I$,  is given by the extended Hubbard term on the kagome lattice
\beq
\mathscr{H}_I &=& \sum_{i j \alpha \beta } U_{j,\alpha \beta}n_{i \alpha }n_{i+j \beta } \quad .
\label{KagomeInteraction}
\eeq
It contains the onsite Hubbard interaction $U$, and nearest and next-nearest neighbor density interactions $V$ and $V'$ respectively. For convenience, these parameters are succinctly expressed in terms of  $U_{j,\alpha \beta}$ and the corresponding number operators are denoted by
$n_{i \alpha}$.
We use the primitive vectors of the triangular lattice  $\bs n_1 = 2 a (\frac{1}{2}, \frac{\sqrt{3}}{2})$, $\bs n_2 = 2 a(-\frac{1}{2}, \frac{\sqrt{3}}{2})$, $\bs n_3 = 2 a(1,0)$ where $a$ is the nearest neighbor inter atomic spacing, to write the minimal band structure from the kinetic Hamiltonian Eq.~\ref{KagomeSOC}. The non-interacting bands 
\qs{of the minimal model}
are shown in Fig.~\ref{kagomebands} (right panel). There are three pairs of bands separated in energy and each is split by the Zeeman field. Each pair has chern numbers $C= \pm 1, 0, \pm 1$ from highest to lowest energy. Note that the total chern number for each pair of bands sums to zero even the presence of the Zeeman field~\cite{FootNote}.  \par
To mimic the 
\qs{realistic electronic structure}
 of FeGe, we place the chemical potential in the lowest pair of flat bands with $C = \pm 1$ as highlighted in the blue region in Fig.~\ref{kagomebands} (right panel). We seek a low energy model describing this pair of bands. To this end, we wannierize the bands to derive exponentially localized WOs~\cite{Vanderbilt2011, Vanderbilt2012, Setty2021}. 
 Denoting $b_{i\mu}^{\dagger}$ as the creation operator in the WO basis on site $i$, the low energy (IR) Hamiltonian obtained after projecting on the  bands crossing E$_F$ is written as $H = H_0 + H_I  $ where
\beq 
H_0 &=& 
\sum_{ij} \sum_{\mu \nu}
b_{i \mu}^\dag t_{ij}^{\mu \nu} b_{j \nu} 
\nonumber
 \\ 
 H_{I} &=& 
 \sum_{ijkl} \sum_{mu\mu'\nu\nu'}
  u_{\mu\mu'\nu\nu'}(j,k,l) b_{i\mu}^\dag b_{i+j\mu'}
 b_{i+k\nu}^\dag b_{i+l\nu'}  \quad .\label{eq:hamiltonian_eff}
\eeq
Here $t_{ij}^{\mu \nu}$ and $u_{\mu\mu'\nu\nu'}(j,k,l)$ are respectively 
the hopping parameters \setty{(see SM Table~\ref{tab:hopping} and SM Fig.~\ref{fig:tbeff})} and interaction matrix elements.  
The interaction parameters are listed in the SM (Tables I-III) of Ref.~\cite{Setty2021}.  
\setty{It is noteworthy that the interaction parameters $u_{\mu\mu'\nu\nu'}(j,k,l)$ are strictly non-local with no onsite interactions.  This is because the wannierization procedure splits the WO centers in real space (See Fig.1(b) of \cite{Setty2021} ). This lack of a common wannier center arises from the $C = \pm 1$ chern distribution and is hence an intrinsic topological property. A key consequence of the wannier center split is the coupling of charge and spin degrees of freedom and is ultimately responsible for the enhanced $T$-symmetry breaking upon entry into the charge order state~\cite{Setty2021}. }\par
\begin{figure}[t!]
\includegraphics[width=1.68in,height=1.25in]{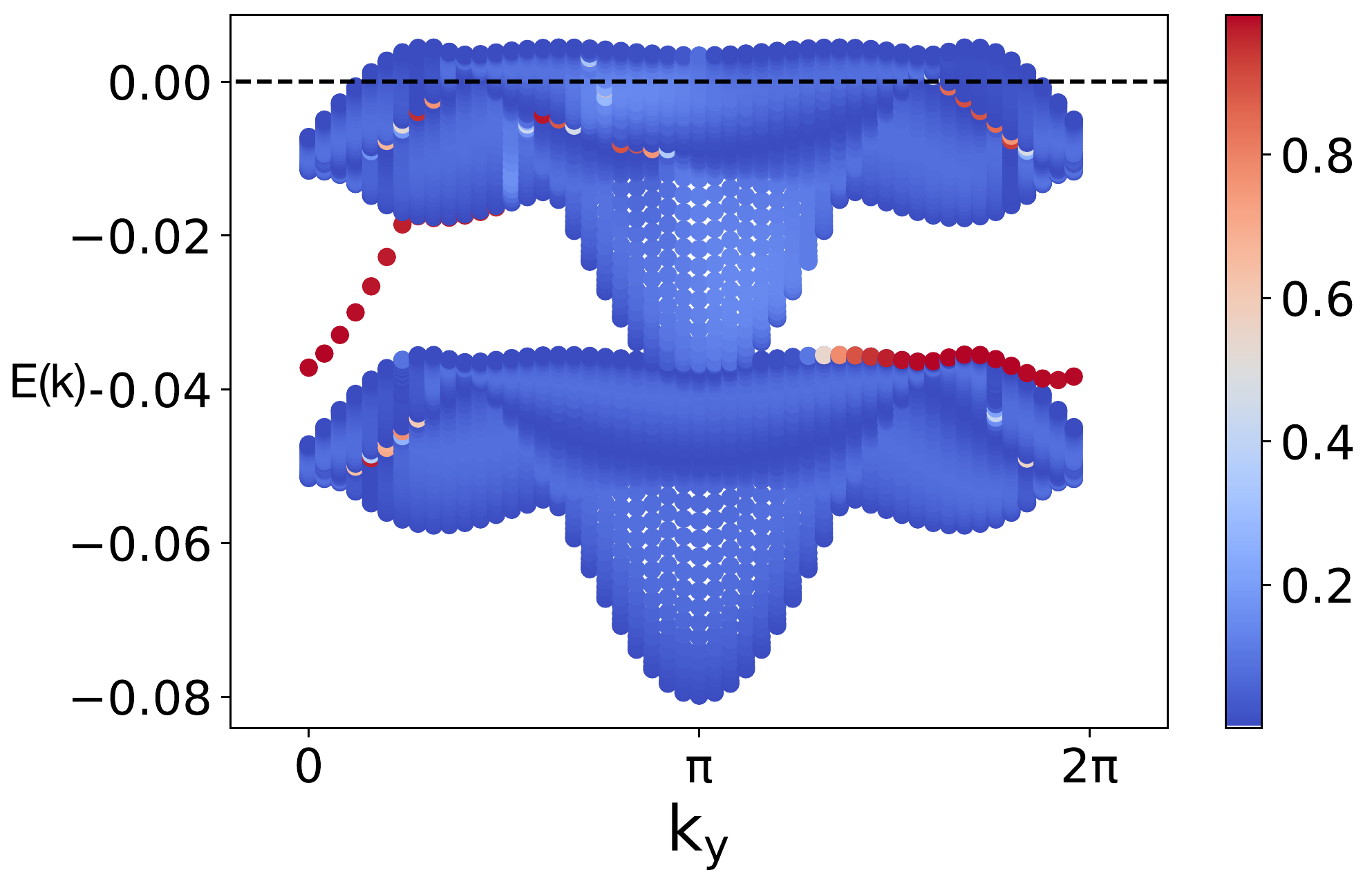}
\includegraphics[width=1.68in,height=1.25in]{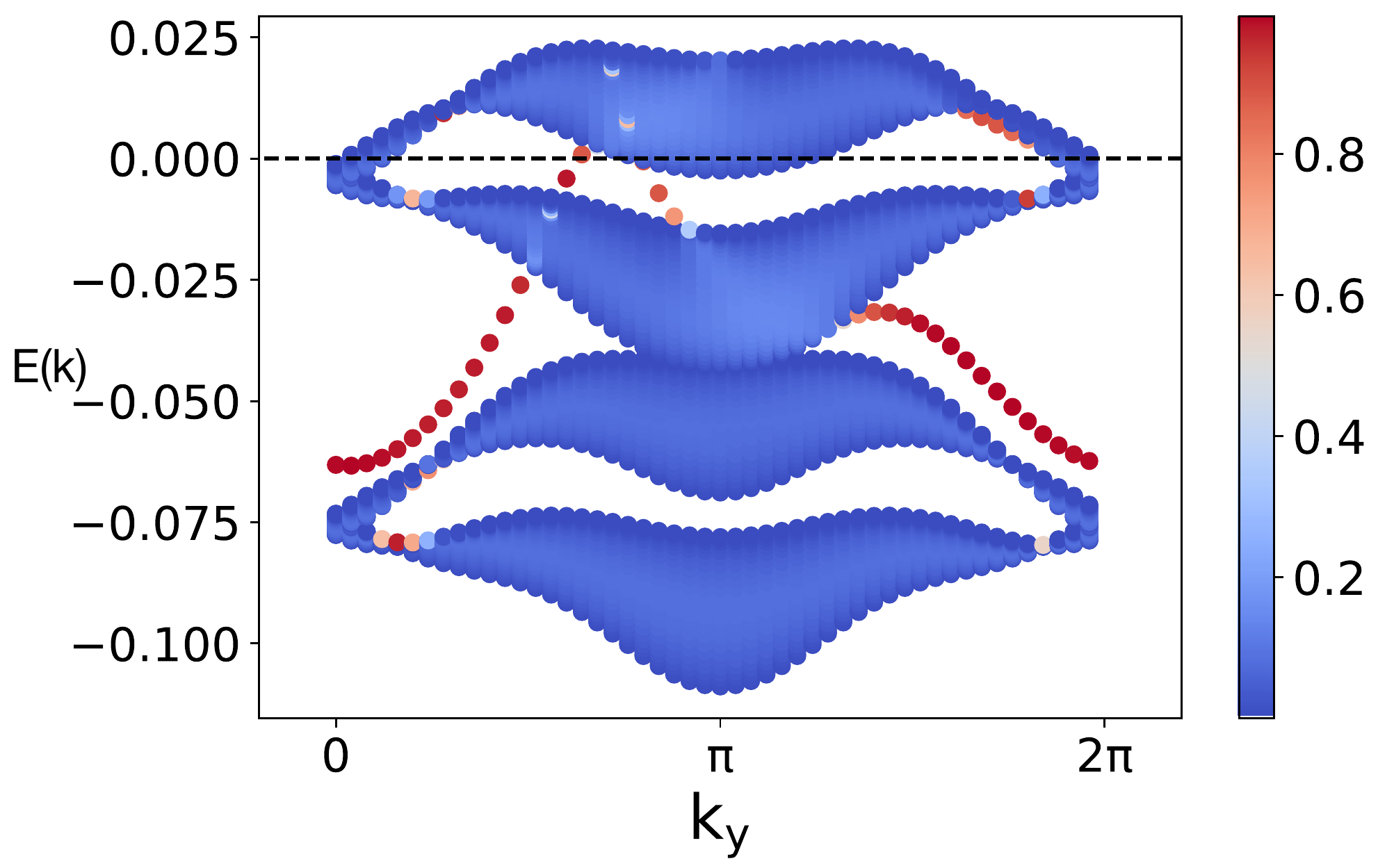}
\caption{ Edge states with the chemical potential located in the upper sub-band and a gapless bulk spectrum. The color scale marks the spectral weight of the states.  (Left panel) Edge states in the parent magnetic order. (Right panel) Edge states in the charge order phase derived from magnetic order.  }\label{edgestates}
%\vskip -0.3 cm
\end{figure}
A full analysis of the Hamiltonian in Eq.~\ref{eq:hamiltonian_eff} without the magnetic order appears in Ref.~\cite{Setty2021}. We have explicitly checked that a similar analysis in the presence of a Zeeman term due to the magnetic order 
yields an analogous phase diagram with the phase boundaries modified only quantitatively.  Above a critical next nearest neighbor interaction 
 $V'$,  a 
 correlation driven charge order phase at the $M$ point in the Brillouin zone (BZ) is stable.  \textcolor{black}{ This order is interesting since, for trivial bands, the two WOs share the same Wannier center and a minimization of the onsite Coulomb repulsion would have yielded a  spin order. Instead, 
in
\qs{our case,}
the non-local $V'$ interaction dominates and drives a charge order at $M$}. 
\par
The charge order also couples to an applied perpendicular magnetic field due to the topology driven mixing between the charge and spin order consistent with recent STM data~\cite{Yin2022}. \setty{For further details regarding these conclusions, we refer the reader to Ref.~\cite{Setty2021}. 
 For the purposes of the current work, we focus on the edge states and enhanced moment in the charge order phase. Here, the Hamiltonian in the $c_{\bs k \alpha \sigma}$ basis takes the form}
 \textcolor{black}{
 \beq \nonumber
H &=& \sum_{\bs k,\alpha,\gamma} \bigg[ c_{\bs k,\alpha,\uparrow}^\dag [h_{0,\bs k}]_{\alpha\gamma} c_{\bs k,\gamma,\uparrow}
 + c_{\bs k,\alpha,\downarrow}^\dag [h^*_{0,-\bs k}]_{\alpha\gamma} c_{\bs k,\gamma,\downarrow}
 \bigg]  \\ \label{MFHamiltonian}
&+& \sum_{\bs k,\alpha} h_0  (c_{\bs k,\alpha,\uparrow}^\dag c_{\bs k,\alpha,\uparrow} 
-c_{\bs k,\alpha,\downarrow}^\dag c_{\bs k,\alpha,\downarrow} ) \\ \nonumber 
&+&\sum_{\bs k,\alpha,\gamma,\sigma,\sigma'} \bigg[\sum_a\phi_{\bs Q}(u_{\bs k+\bs Q,\alpha,\sigma}^a)^*u_{\bs k,\gamma,\sigma'}^a\bigg] c^\dag_{\bs k+\bs Q,\alpha,\sigma} c_{\bs k,\gamma,\sigma'} 
\eeq
where $\alpha,\gamma$ denotes sublattice indices.
$h_{0,\bs k}$ is the hopping matrix of spin up electron with bare dispersions $\epsilon_{\bs k}, \epsilon_{\bs k + \bs Q}$ and $h_0$ is the magnetic field. $\phi_{\bs Q}$ denotes the charge order parameter in the wannier-orbital basis which is a real number. $u_{\bs k,\alpha,\sigma}^a$ is the wavefunction of WO $a$, defined as
 $b_{a,\bs k} =\sum_{\alpha,\sigma} u_{\bs k,\alpha,\sigma}  c_{\bs k,\alpha,\sigma}$, and is related to the Haldane matrix as described in Ref.~\cite{Setty2021}.} \par \setty{
The terms that mix two opposite spins $\sigma$ and $\bar{\sigma}$
 \qs{(Eq.~\ref{MFHamiltonian}, with a finite momentum $\bs Q$)}
 couple charge and spin.
They would not have existed
  for charge order 
  from topologically trivial bands when the SOC is directed out of the kagome plane. 
  The uniqueness of such terms to correlated orders derived from topological bands is due to wannier center splitting  discussed above~\cite{Setty2021}. This effect will be crucial for understanding the charge order-induced magnetic moment appearing later. } 
  \par
  \textcolor{black}{We can now obtain edge states by solving Eq.~\ref{MFHamiltonian} on a slab.  Fig.~\ref{edgestates} shows a plot of the edge state spectrum along the $x=0$ edge in the reduced BZ. The spectral intensities are shown in a color scale. The left (right) panel corresponds to the edge state spectrum in the parent magnetic order (charge order) phase.   Whereas the edge states connecting the top and bottom sub-bands are overwhelmed by the gapless bulk bands in the magnetic parent phase, they completely dominate in the charge order phase.  Additional edge state analysis when the bulk is fully gapped by the Zeeman field 
  appears in the SM Fig.~\ref{edgestates-gapped}.   The origin of the edge state behavior is tied to the non-zero chern numbers of the kagome flat bands crossing E$_F$ and a non-trivial coupling between the magnetic and charge orders. The out-of-plane field of the 
magnetic
  order splits the non-interacting bands with equal and opposite chern numbers and the low energy (non-interacting) model written in the original $c$ electron basis can be mapped to the Haldane bands. Hence the suppression of the density of states at the chemical potential due to the charge order enables the observation of edge states as observed in STM~\cite{Yin2022}. } \par
\textcolor{black}{We now illustrate our topological mechanism for the enhanced
magnetic moment,
\qs{at the in-plane $\bs q=0$,}
along the positive $c$-axis in the charge ordered phase.} 
 We do this by expanding Eq.~\ref{MFHamiltonian} in a Ginzburg-Landau type of  free energy in terms of the charge order parameters $\Delta_{\bs Q} \sim \sum_{\bs k} \langle c_{\bs k \alpha \uparrow}^{\dagger} c_{\bs k+ \bs Q \alpha \downarrow} \rangle $ and $\Delta'_{\bs Q} \sim \sum_{\bs k} \langle c_{\bs k + \bs Q \alpha \uparrow}^{\dagger} c_{\bs k \alpha \downarrow}\rangle $.  For a single sub-lattice $\alpha$, this is given as
\beq 
\label{FreeEnergy} 
F[\Delta_{\bs Q_i}, \Delta_{\bs Q_i}'] &=& F_0[\Delta_{\bs Q_i}, \Delta_{\bs Q_i}'] + \Delta F[\Delta_{\bs Q_i}, \Delta_{\bs Q_i}'] \\ \nonumber
\Delta F[\Delta_{\bs Q_i}, \Delta_{\bs Q_i}'] &=& - h' \gamma \sum_i \left(\Delta_{\bs Q_i} \Delta_{-\bs Q_i} + \Delta_{\bs Q_i}' \Delta_{-\bs Q_i}' \right). 
\eeq
Here $F_0$ is independent of a `probe' field $h'$ 
\qs{(SM, Eq.~\ref{FullFreeEnergy}),}
 and $\gamma \simeq \frac{3 h_0}{2 T_{CO}^4} \left[ \frac{-1+ \tilde h_0^{-1}\sinh \tilde h_0}{\tilde h_0^2 (1+ \cosh \tilde h_0)} \right]$ with $\tilde h_0 = h_0/T_{CO} $ in the nested, flat band limit of $\epsilon_{\bs k}=\epsilon_{\bs k+ \bs Q_i} \sim 0$.  $\bs Q_i$ are the $M$ point wave vectors corresponding to the $2 \times 2$ charge order. The  $\Delta F$ term (see SM Fig.~\ref{FD}) couples charge and $\bs Q=0$ magnetic orders and is odd in the static and probe fields $h_0$ and $h'$. 
The induced magnetic moment becomes  
 \beq
\delta M = - \frac{\partial F[\Delta_{\bs Q_i}]}{\partial h'} \propto \gamma (\Delta_{\bs Q_i}\Delta_{-\bs Q_i}+\Delta'_{\bs Q_i}\Delta'_{-\bs Q_i}). \label{dm}
\eeq 
    From the expression of $\gamma$ above, the induced $\delta M$ is \textit{along} the direction of the static magnetic field ($h_0$) and rises quadratically in the order parameter upon entry into the charge order phase. This result is consistent with neutron scattering data~\cite{Dai2022}.
 \par
    \setty{As alluded to earlier, the terms proportional to $\gamma$ in Eqs.~\ref{FreeEnergy}, \ref{dm} have a topological origin. From the definitions of  $\Delta_{\bs Q}$ and $\Delta_{\bs Q}'$, it is evident that the order parameters have a finite wave vector and mix opposite spins, hence coupling spin and charge. } 
This is absent for charge order derived from trivial bands with an out-of-plane SOC.  In fact, the finite wave vector non-spin flip terms of the type  $ \sum_{\bs k} \langle c_{\bs k + \bs Q \alpha \sigma}^{\dagger} c_{\bs k \alpha \sigma}\rangle $ do not pin the induced magnetic moment to $h_0$.  Hence the induced $\delta M$ \textit{along} the static $h_0$ is a fingerprint of topology of the parent magnetic phase.   \par
\textit{Discussion:}
 Several points are in order. 
  While the suppression of the DOS close to the chemical potential due to the charge order enables the observation of edge states in STM, it should be noted that the bulk states need not be fully gapped or even nodal for the observation of edge states. 
  Even
  a modest suppression of the spectral weight at  E$_F$ -- 30 \%- 40\% of the high temperature value with a large residual density of states -- is sufficient for the edge spectrum to be discernible. This is consistent with our calculations in Figs~\ref{edgestates} where edge states have a large spectral weight despite the lack of a full gap in the charge order phase.   \par In principle, such topological edge states
 should exist in the purely magnetic phase as well ($T_{CO}$$<T<T_m$). However,  both 
  DFT and DMFT
demonstrate
large density of states of the bulk bands near E$_F$
in this regime. Correspondingly, edge
states 
 are
 expected to be 
 \qs{smeared} by the
 bulk bands crossing E$_F$.
 This picture is supported by our calculations in Fig.~\ref{edgestates}.
In fact, topological surface states are expected to exist even in the normal state above $T_m$ due to the non-trivial topology of the bands, 
albeit mixed with the more dominant bulk states. 
Recently such states were observed in photoemission measurements of the ternary kagome compounds~\cite{Shi2021} 
in both the normal and charge ordered phases.  
\par
We have delineated the role of van Hove singularities in the electronic structure. 
They are located $\sim -0.6eV$ below E$_F$ in the 
 nonmagnetic phase, and are pushed closer to E$_F$ 
in the magnetic phase. However,
the flat bands 
always dominate the DOS.
 As seen in the bottom panels
Fig.~\ref{ElectronicStructure}, and further highlighted in SM (Fig.~\ref{vHove}),
the density of states (DOS) from the van Hove singularities is essentially
completely obscured by the 
topological
flat bands.
This
  renders the possibility of a strong coupling topological mechanism for the charge order more viable and robust~\cite{Setty2021}.
  Finally, 
  we have illustrated the topological mechanism by focusing on an effective model with only intra-layer coupling.
  Generalization of our analysis to \setty{the more realistic} 3D case is 
   left for
  future investigations.
  \par 
 \textcolor{black}{In conclusion, the recent discovery of charge order in the topological flat band magnetic kagome metal FeGe opens up a new platform for exploring flat band correlated phases and their topological properties. } While the charge order in FeGe shares many features with its ternary counterparts $A$V$_3$Sb$_5$ ($A$ = K, Cs, Rb), it is also distinct in 
 important ways. The strong correlation
 driven band renormalizations, active topological flat bands, appearance of edge states, 
 and enhancement of the 
 magnetic
 moments below the charge ordering temperature are a few key aspects unique to FeGe. 
 \setty{Beyond
 what was discussed in the nonmagnetic setting \cite{Setty2021},
 namely
 correlation driven charge order,
 ordering at the
 wavevector
  $M$ in the BZ,
  and 
  breaking of 
  time-reversal symmetry,
  } 
  we have shown here that our minimal model in a magnetic setting provides a
  qualitative understanding of
  \qs{the most}
   salient and puzzling features of the charge order in FeGe:
  (i) the existence of edge states 
  and  (ii) 
  development of an excess
  magnetic
  moment with 
  the in-plane
  \qs{ $\bs q=0$.}
The origin of all these exotic properties can be  traced to the non-trivial band topology of the magnetic parent phase. 
 As such, our results will likely have general implications on the interplay of topology and correlations in a broad range of correlated systems.
 \par
 \textit{Acknowledgements:} We thank P. C. Dai, M. Yi, J. X. Yin for
sharing their results with us and for useful discussions. 
This work has in part been supported by the
AFOSR under Grant No.\ FA9550-21-1-0356 (C.S. and Q.S.)
and the Robert A. Welch Foundation Grant No.\ C-1411 (L. C. and H.H.).
Work at Los Alamos was supported by LANL LDRD Program, UC Laboratory Fees Research Program (Grant Number: LFR-20-653926), 
 and in part by Center for Integrated Nanotechnologies, a U.S. DOE BES user facility.
 One of us (Q.S.) acknowledges the hospitality of the Aspen Center for Physics,
which is supported by NSF grant No. PHY-1607611.
  \\ \newline
 \noindent
$\oplus$ These authors contributed equally to this work.\\
$\dagger$ Email: csetty@rice.edu \\
\bibliography{FeGe.bib} 
\newpage
\onecolumngrid
\newpage
\section{Supplemental Material}
In this Supplemental Material, we further expand on the computational aspects of the electronic structure and edge state analysis. \\ \par
\begin{figure}[h!]
\includegraphics[width=2.2in,height=1.75in]{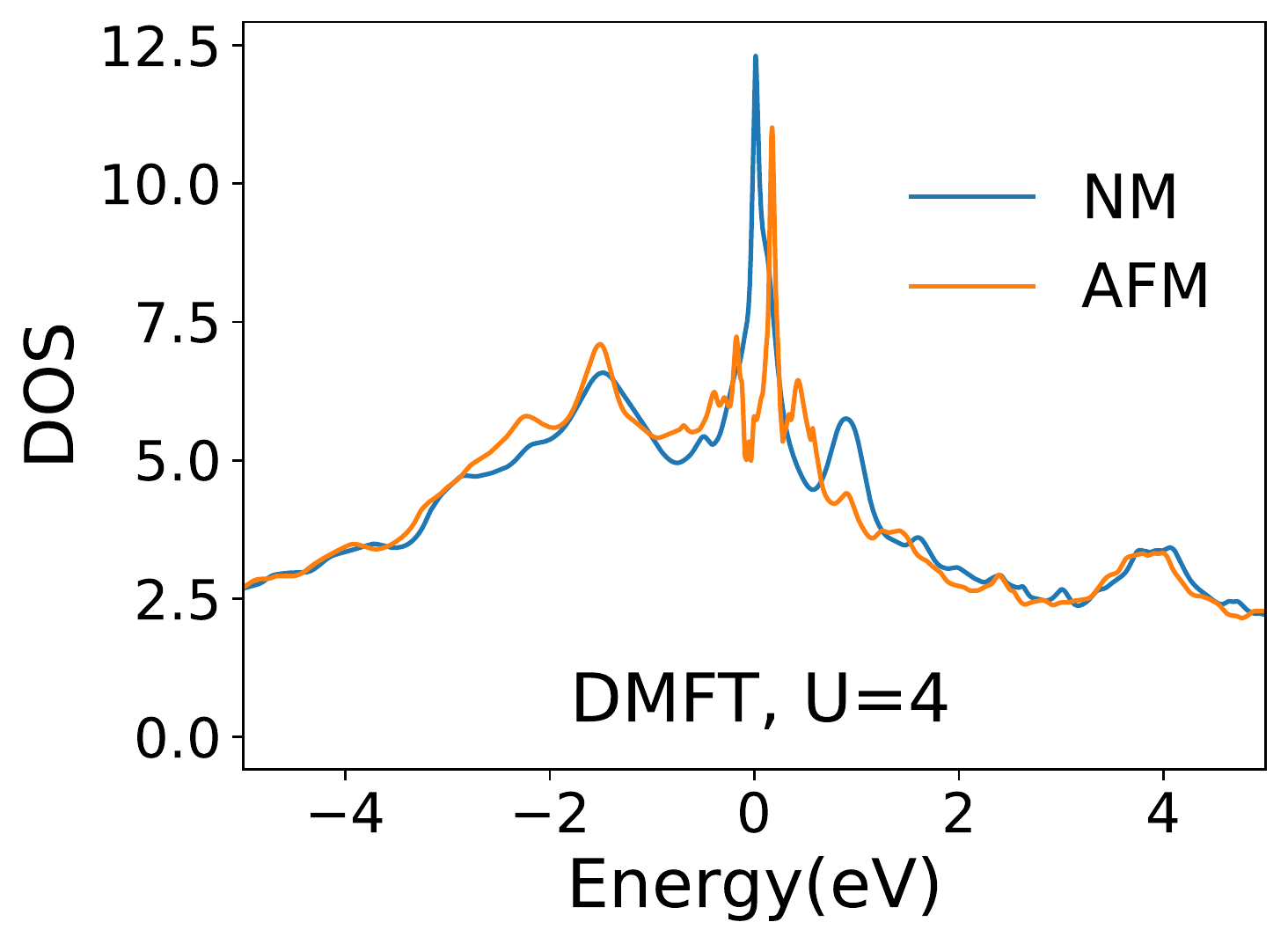}
\includegraphics[width=2.2in,height=1.75in]{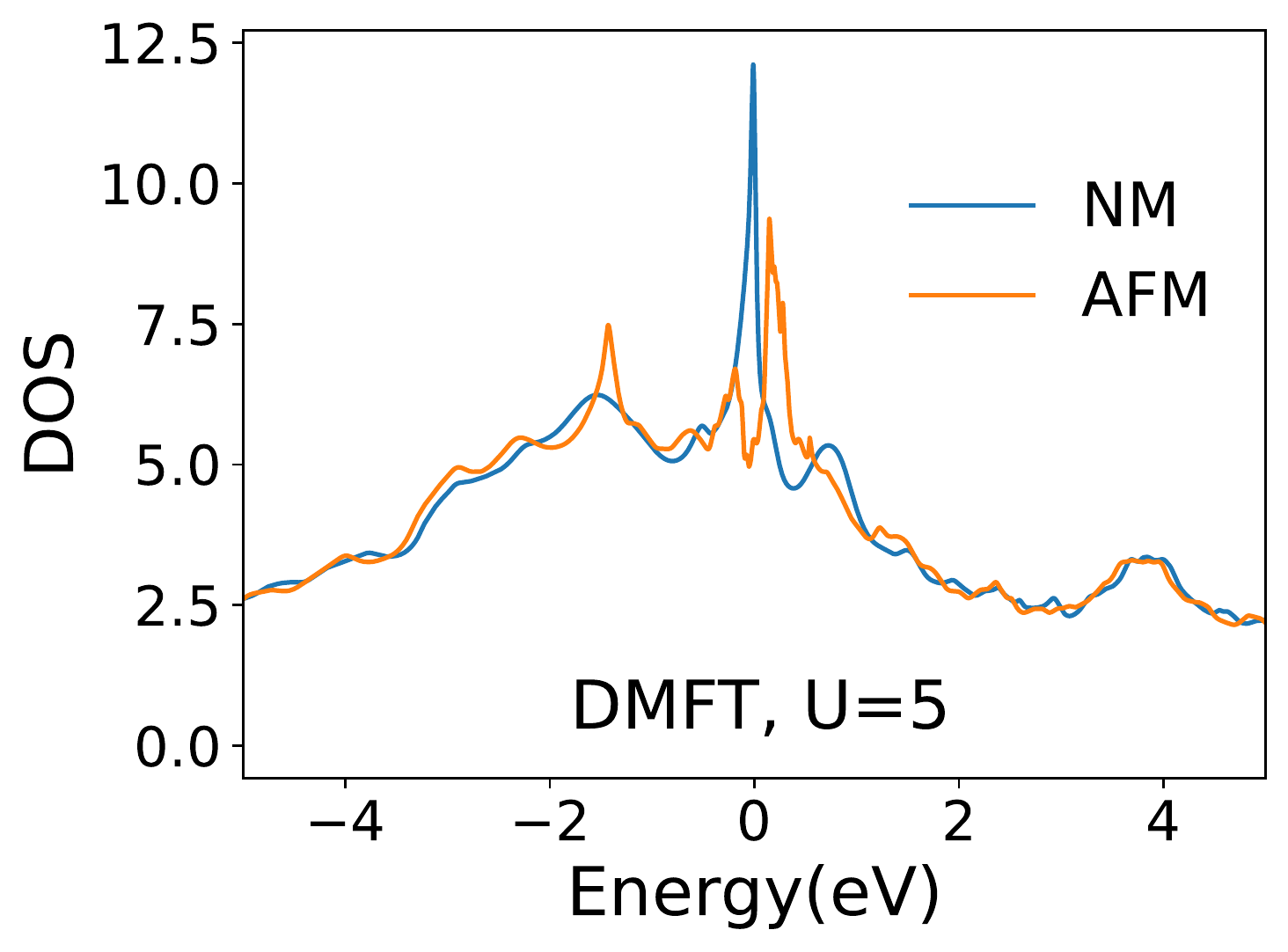}
\includegraphics[width=2.2in,height=1.75in]{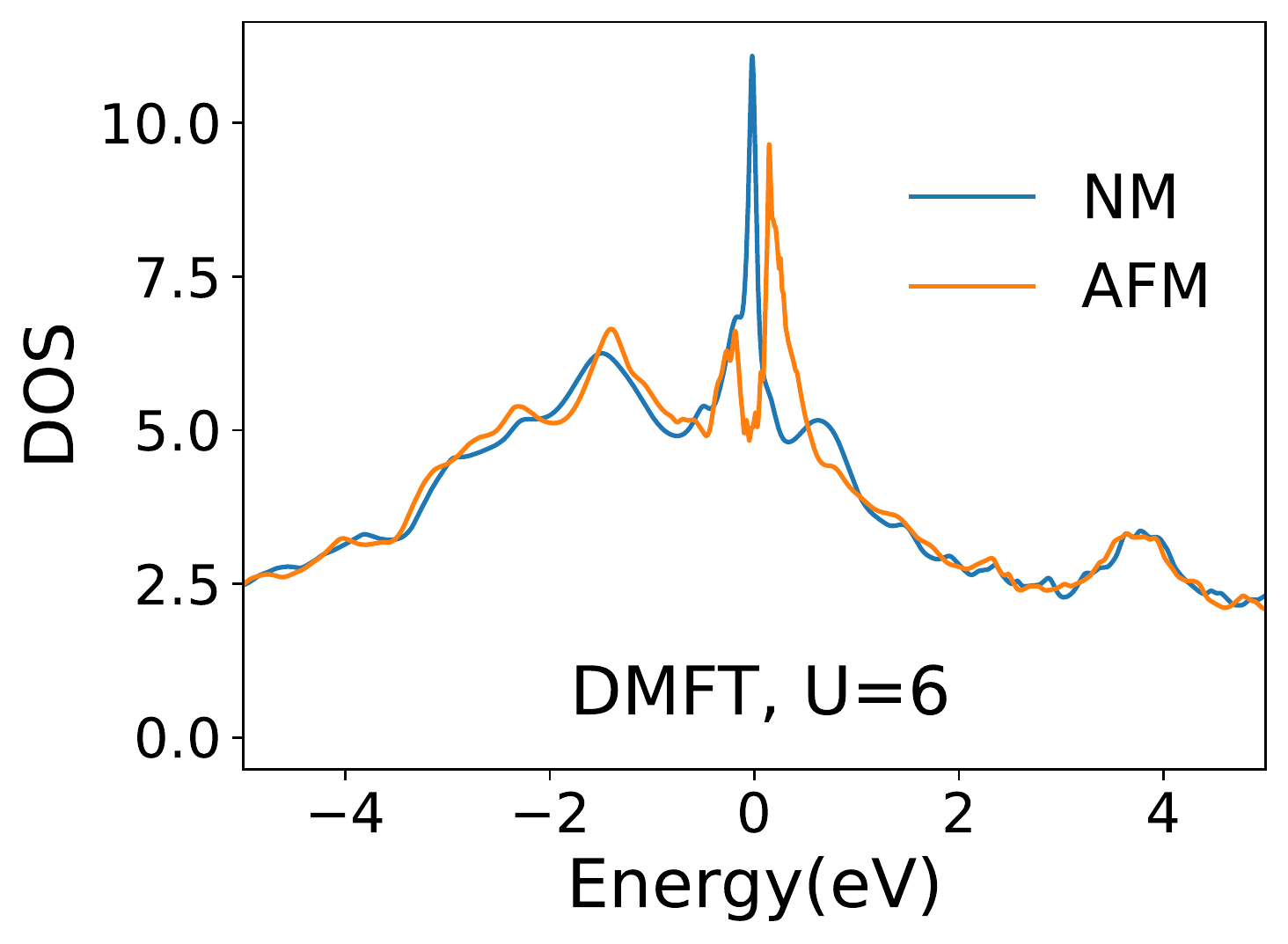}
\includegraphics[width=2.2in,height=1.75in]{FeGe-PMso_U4p0}
\includegraphics[width=2.2in,height=1.75in]{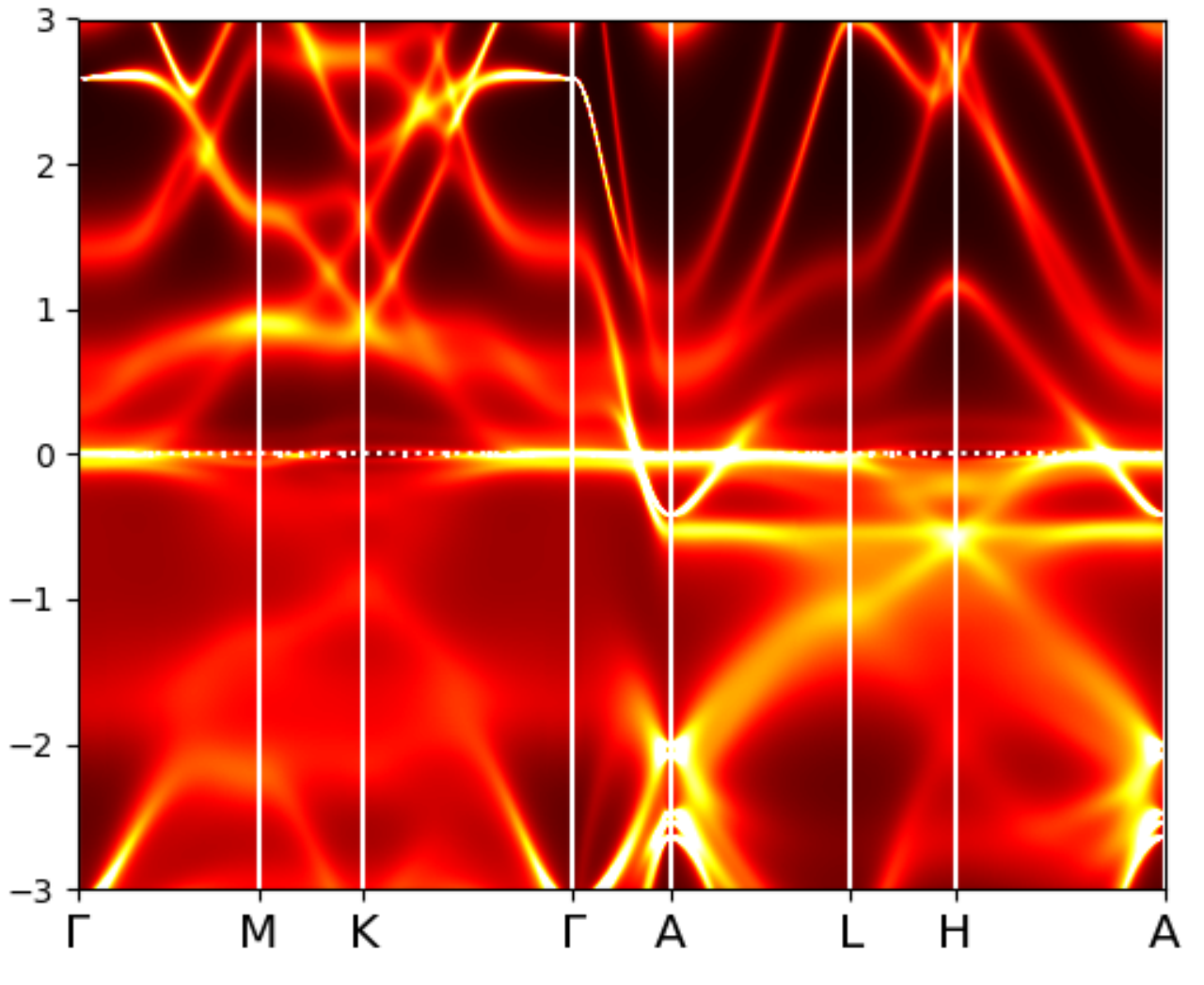}
\includegraphics[width=2.2in,height=1.75in]{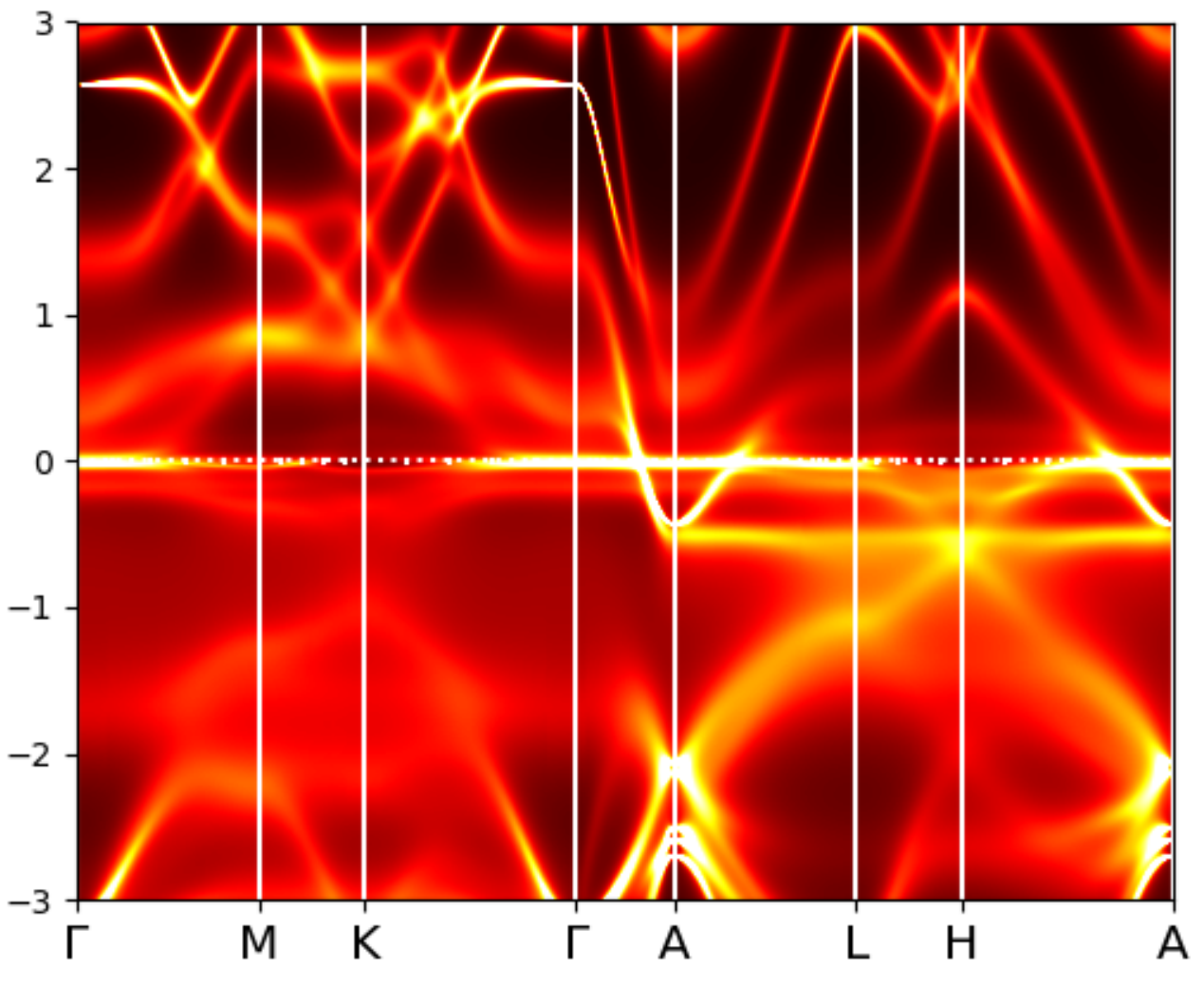}
\includegraphics[width=2.2in,height=1.75in]{FeGe-AFMAso_U4p0}
\includegraphics[width=2.2in,height=1.75in]{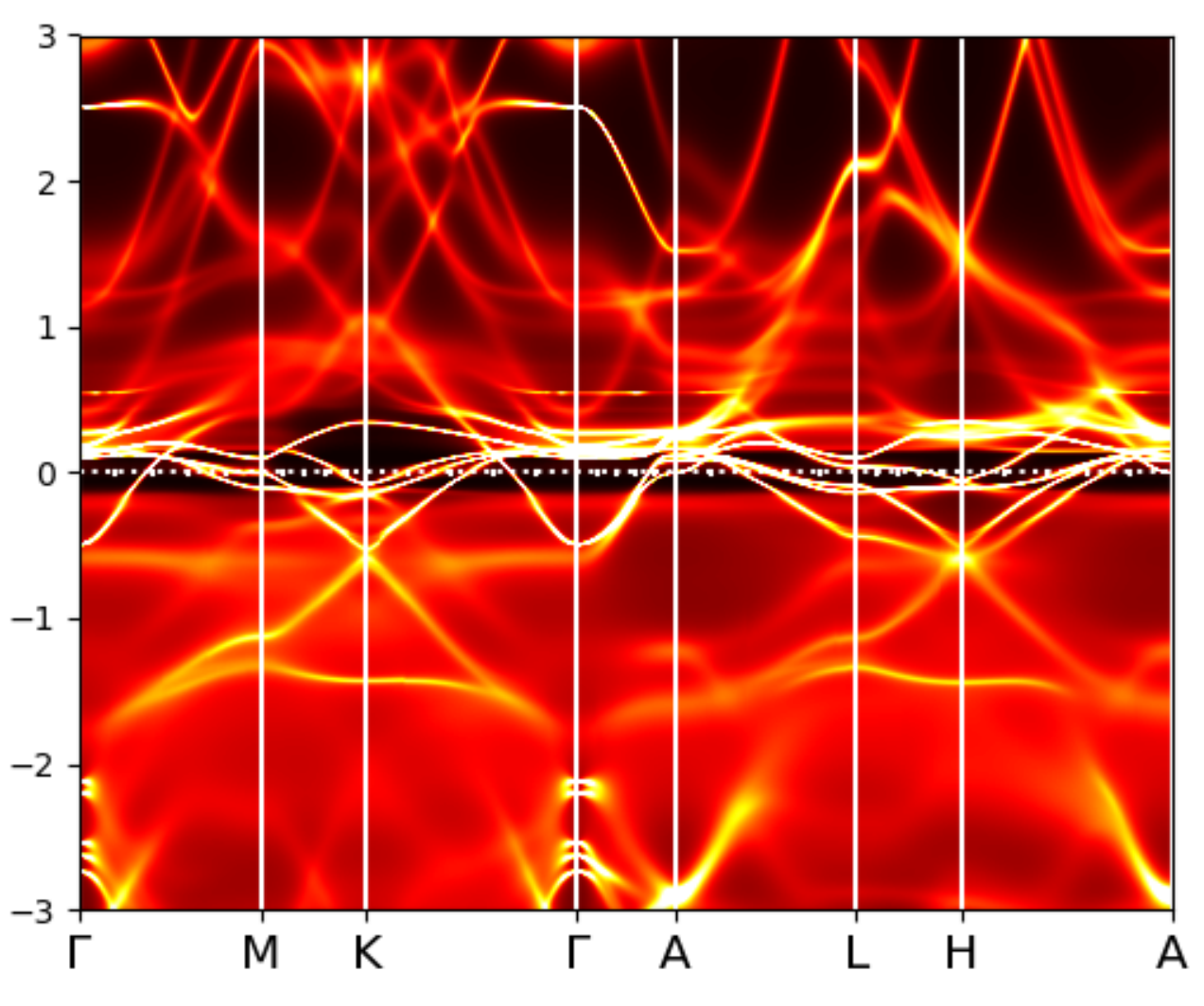}
\includegraphics[width=2.2in,height=1.75in]{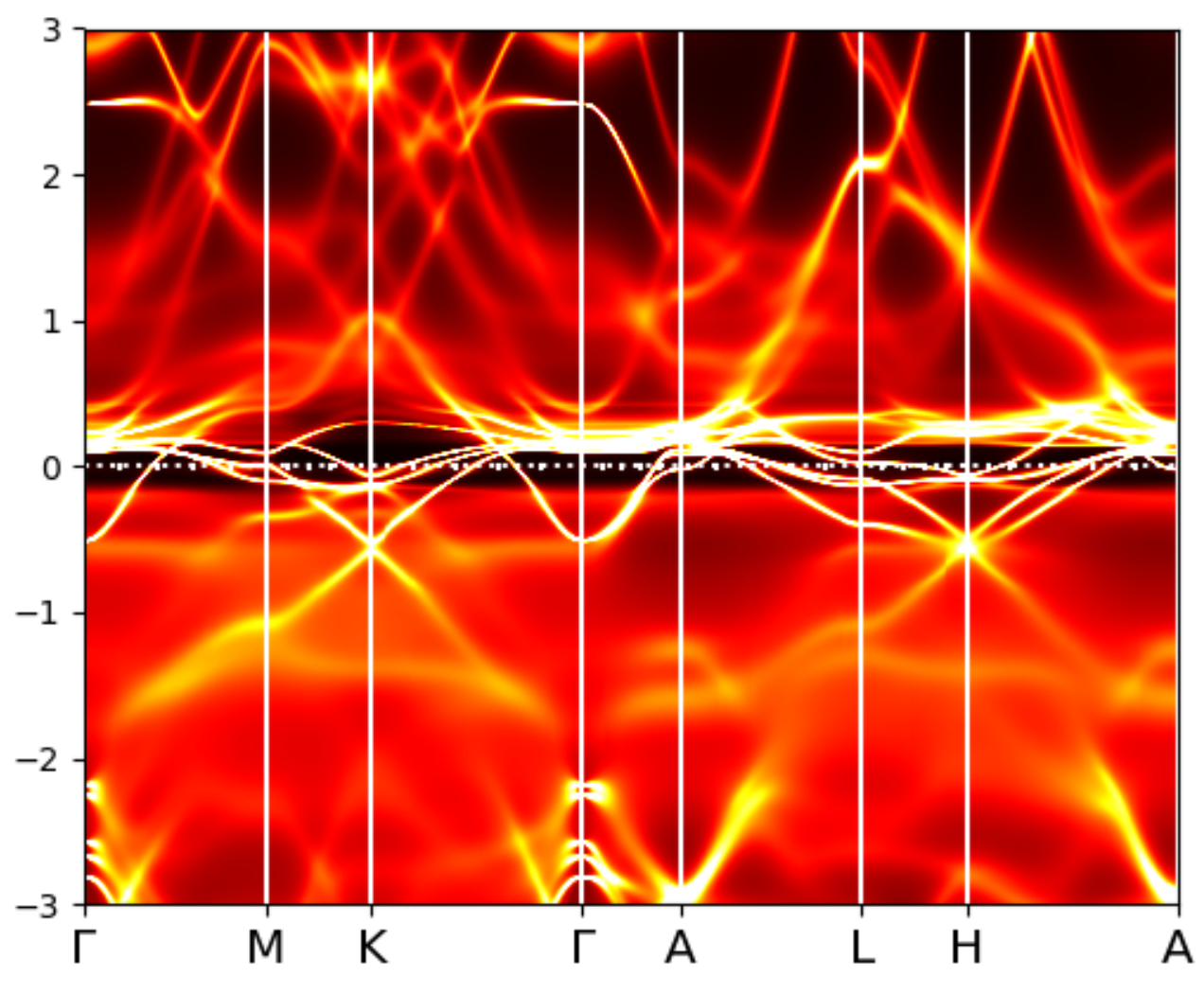}
\caption{ Top row: DOS calculated from DMFT in the NM and magnetic phases. The plots are over a larger energy window than shown in the main text. (Middle row) DMFT bands with spectral intensities in the NM phase with SOC. (Bottom row) DMFT bands with spectral intensities in the magnetic phase with SOC. The columns from left to right correspond to  $U=4 eV, 5 eV, 6 eV$ respectively.
}\label{DOS-Supp}
%\vskip -0.3 cm
\end{figure}

\textit{Role of Coulomb interactions:} Fig.~\ref{DOS-Supp} shows the DMFT DOS (top row) and spectral intensities (middle and bottom rows)  as a function of increasing coulomb interaction $U$ (columns from left to right correspond to $U=4 eV, 5 eV, 6 eV$ respectively). Many of the important low energy features such as location of the flat bands near E$_f$ and Dirac cones remain intact with increasing correlations both with (bottom row) and without magnetic order (middle row). However, both the flat bands and Dirac cones at $M$ become narrower and more localized as is natural when the electrons become strongly correlated. The spectral intensities are also suppressed  indicating loss of quasiparticle-like behavior.  \par
\begin{figure}[h!]
\includegraphics[width=3.5in,height=3in]{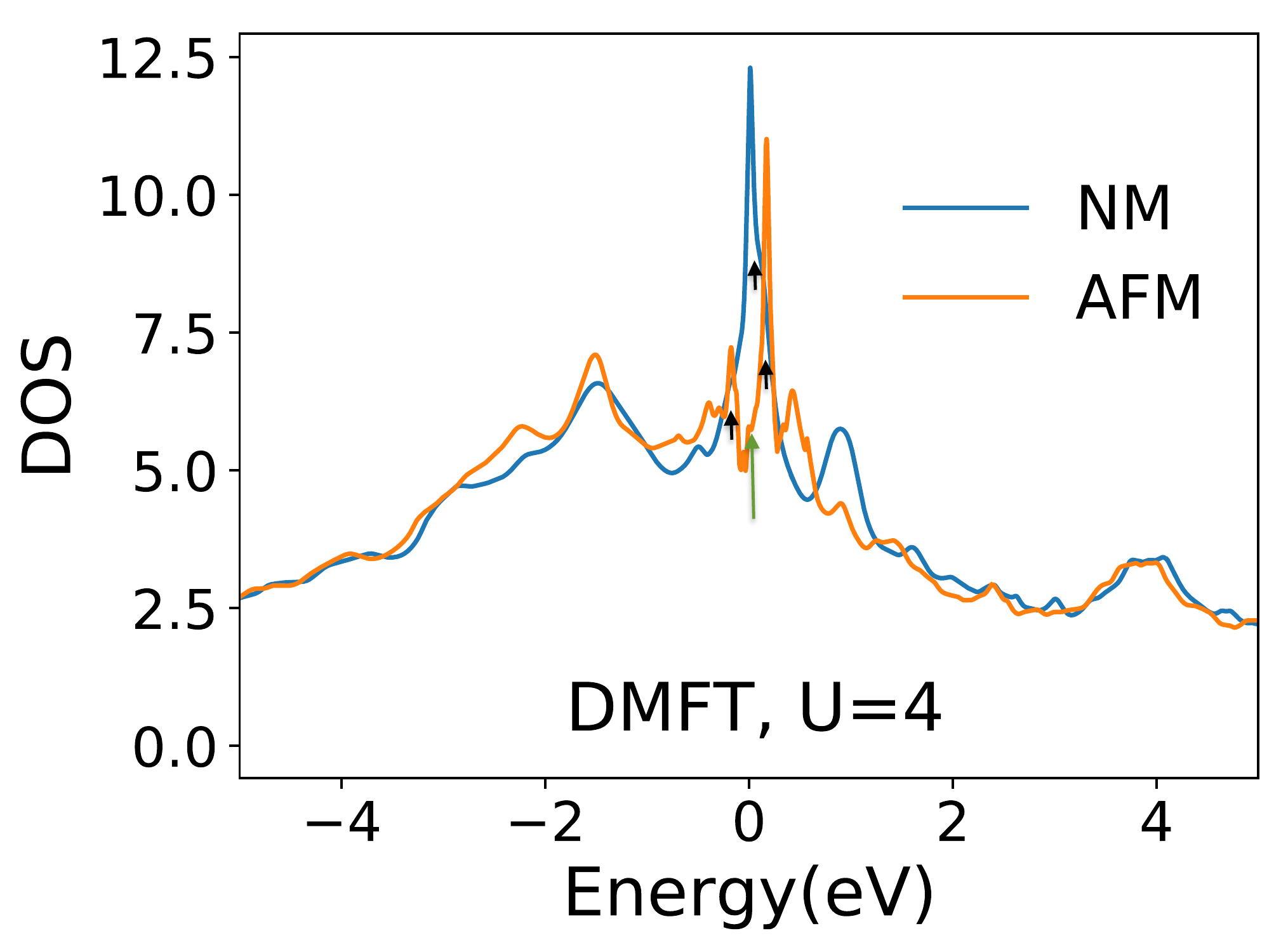}
\includegraphics[width=3.5in,height=3in]{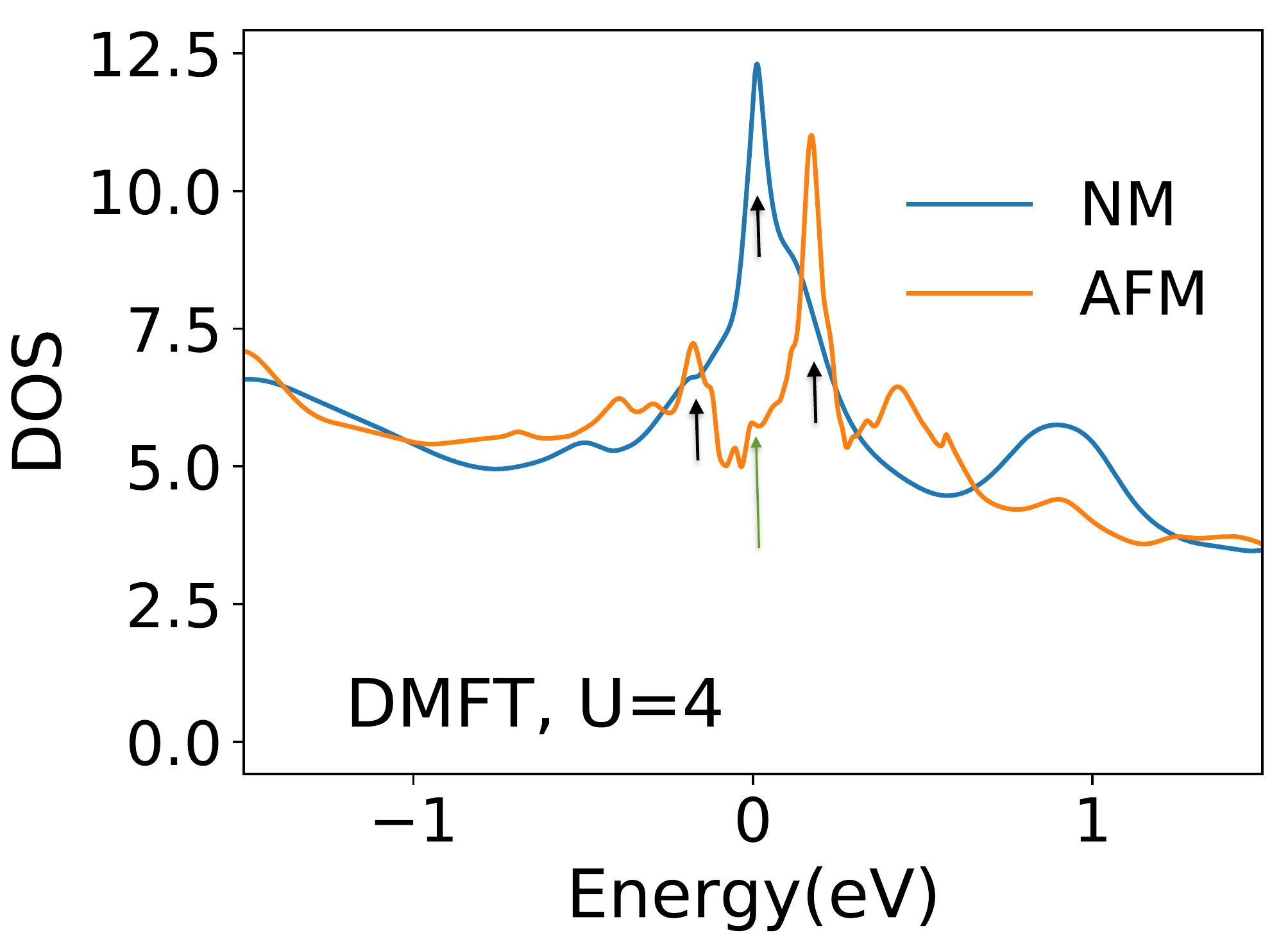}
\caption{ Flat band vs vHove singularity contribution to the DOS in the NM and magnetic phases as calculated in DMFT. The black short arrows denote the dominant flat bands while the green long arrow denotes the v-Hove contribution. The right panel is a zoomed in version of the left for clarity. 
}\label{vHove}
%\vskip -0.3 cm
\end{figure}

\textit{Role of vHove singularities:} Fig.~\ref{vHove} (left) marks different high intensity peaks and satellites features in the DOS and their relationship to various properties of the electronic structure. Fig.~\ref{vHove} (right) shows the same plot zoomed into a narrower energy window. Like in the main text, the short black arrows indicate peaks in the DOS that are determined by the flat bands. The long green arrow  highlights the contribution from the vHove singularity. From a comparison of the relative intensities, it is evident that the role of vHove singularities is obscured by the more dominant flat band effects.  This is unlike the 135 ternary kagome compounds where the flat bands are buried deep below the fermi energy and vHave singularities have a greater influence on the low energy properties.  \par
\begin{figure}[h!]
\includegraphics[width=3.5in,height=3in]{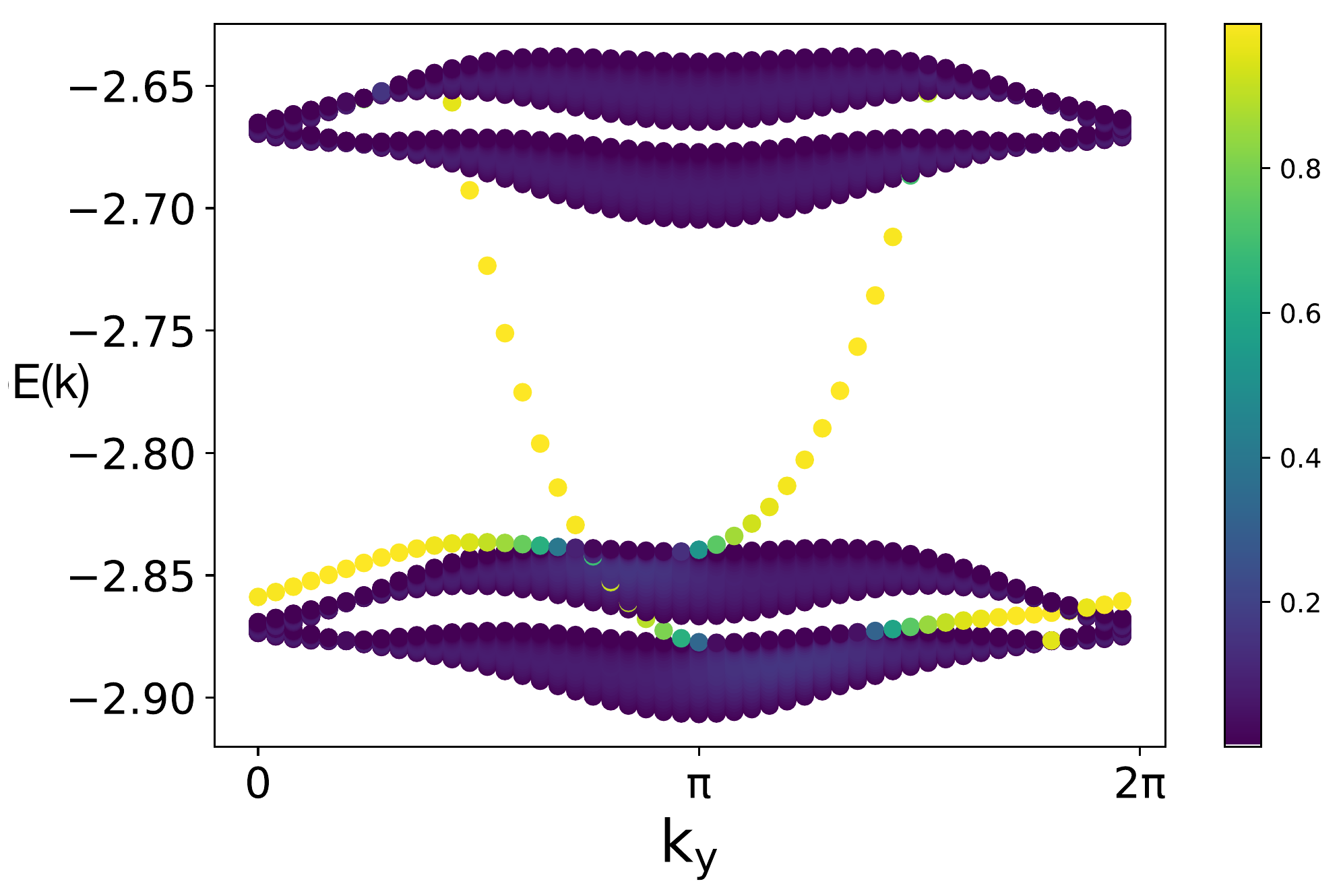}
\includegraphics[width=3.5in,height=3in]{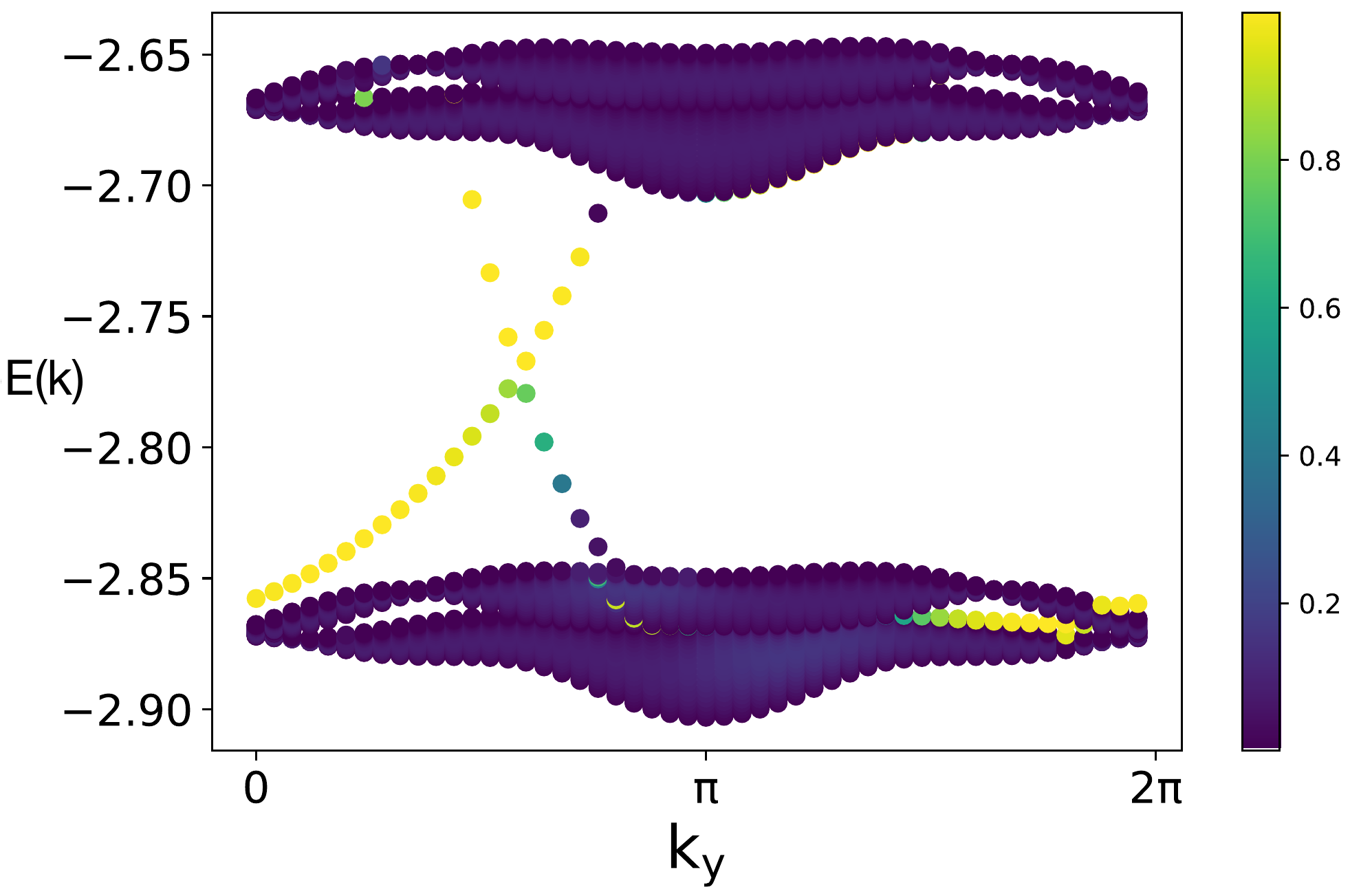}
\caption{ Edge states in the charge order phase when the magnetic order is strong enough to fully gap out the bulk. The color scale marks the spectral weight of the states. In the left panel, the charge order parameter is comparable to the parent magnetic order. In the right panel, the charge order parameter is much smaller than the magnetic order parameter. 
}\label{edgestates-gapped}
\end{figure}
\textit{Large magnetic order:} Here we consider the case where magnetic order completely polarizes the chern bands and gaps out the bulk at half filling. Fig.~\ref{edgestates-gapped} (left) shows the edge states when the charge order parameter is comparable to the parent magnetic order. Edge states persist at both quarter filling and half filling. A similar scenario follows when the charge order parameter is much smaller than the magnetic order parameter (Fig.~\ref{edgestates-gapped} (right)). However, note that in this scenario, the bulk state contribution to the STM DOS is non-existent at half filling. Whereas at quarter filling, there are both bulk and edge state contribution to the DOS.  \\ \newline
\textit{Projected Hamiltonian in the Wannier basis:}
In this section, we derive the effective model in the Wannier basis. As shown in Ref.~\cite{Setty2021}, the lowest two bands of the extended Hubbard model with Coulomb interaction on the kagome lattice can be projected into an effective Hamiltonian defined on the WOs. The projected creation and annihilation operators in the kagome lattice is connected with WOs with a sequence of unitary transformations: 
\begin{equation}
P c_{{\bf k}} P = \sum_{\mu} (V_{{\bf k}}^{\dagger}U_{H, {\bf k}})_{\mu\nu} b_{{\bf k}\nu},
\end{equation} 
where $V_{{\bf k}} $ diagonalizes the hopping matrix, and $U_{H, {\bf k}}^{\dagger}$ is the Haldane matrix~\cite{Vanderbilt2012}. The kinetic part of effective Hamiltonian reads:
\begin{equation}
H_{0} = \sum_{ij,\mu\nu} b^{\dagger}_{i\mu} t_{ij}^{\mu\nu} b_{j\nu},
\end{equation}
where 
\begin{equation}
t^{\mu\nu}_{ij} = \sum_{\mu'\nu'i'j'} T^{\mu'\nu'}_{i'j'} \omega^{*}_{\mu'\mu,i'-i}\omega_{\nu'\nu,j'-j},
\end{equation}
with $\omega_{\alpha,\nu,i}=\frac{1}{N}\left( V_{{\bf k}}^{\dagger}U_{H,{\bf k}}^{\dagger} \right)_{\mu\nu} e^{i{\bf k}\cdot {\bf r_i} }$ the real space wave functions of the WOs. 

\begin{figure}[h]
\includegraphics[angle=0,width=0.5\linewidth]{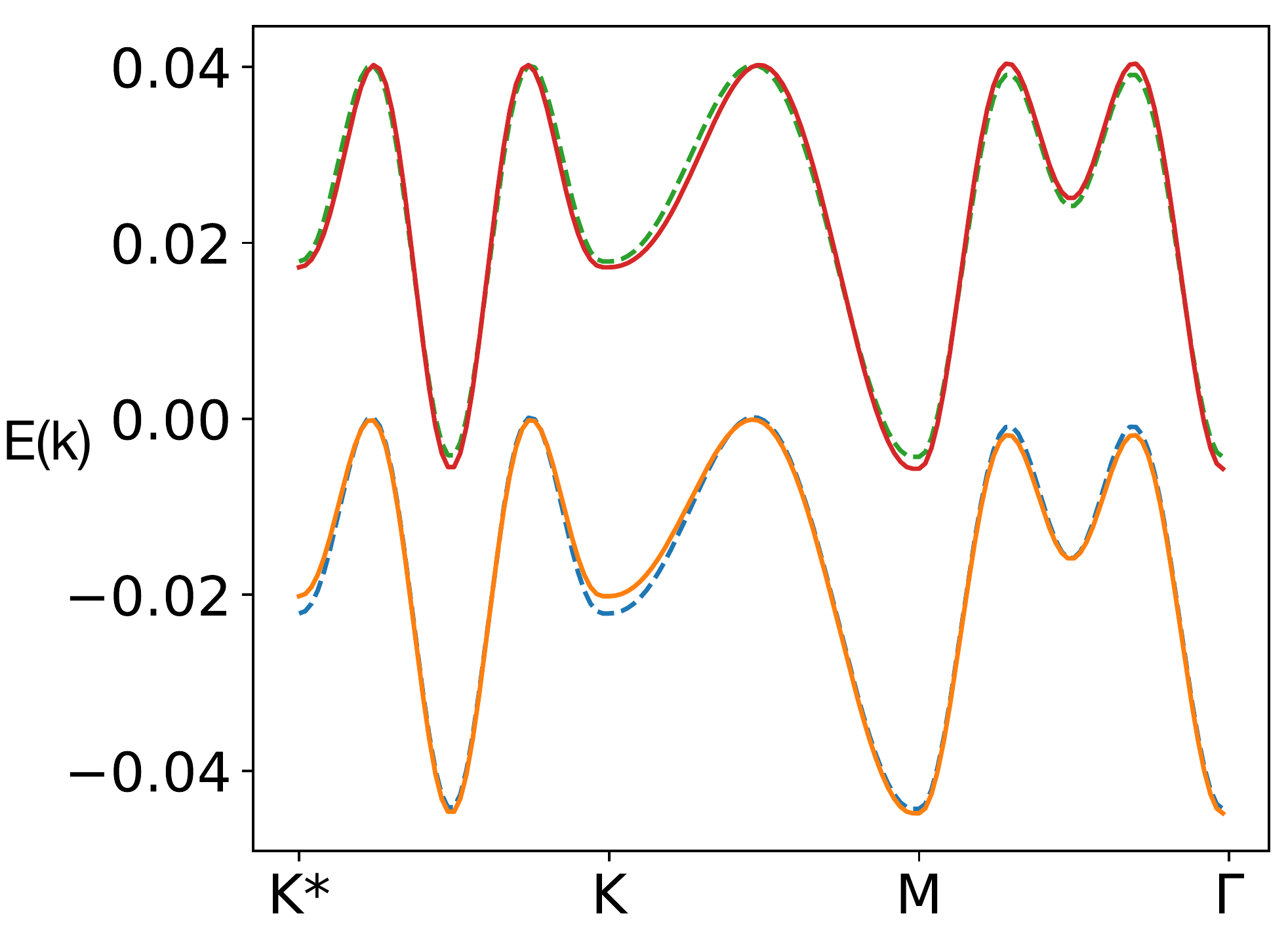}
\caption{ Comparison of dispersions along high symmetry lines of the lowest degenerate doublet of kagome bands. Dashed line
is the original kagome band and solid line is from the effective model with WOs.
}\label{fig:tbeff}
%\vskip -0.3 cm
\end{figure}

\begin{figure}[h!]
\includegraphics[width=3.5in,height=3in]{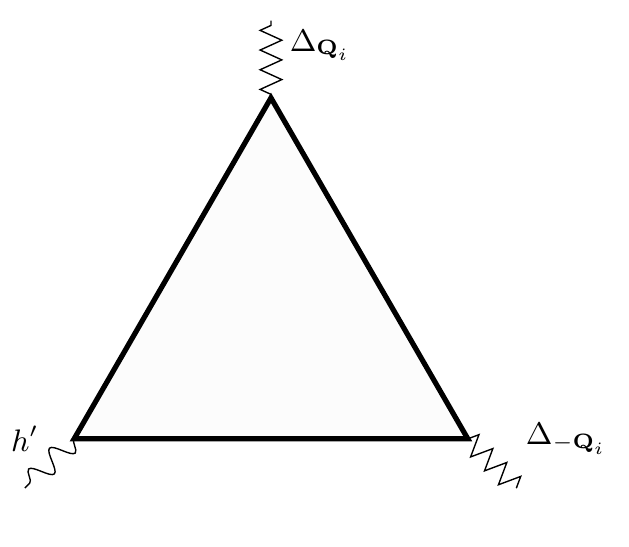}
\includegraphics[width=3.5in,height=3in]{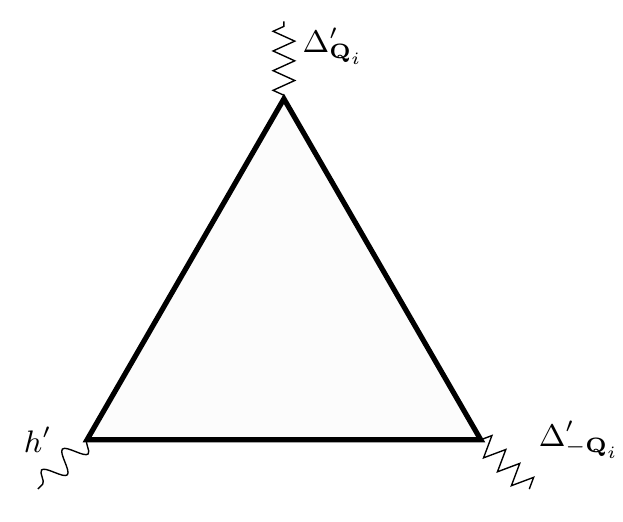}
\caption{ Lowest order Feynman diagrams that couple the charge order fluctuations $\Delta_{\bs Q_i}$ and $\Delta_{\bs Q_i}'$ (zig-zag lines) to fluctuations of the broken time reversal field (wavy lines) $h'$. The solid lines are the electron Green functions in a static external field $h_0$.  These terms drive the enhanced magnetic moment in the charge order phase. 
}\label{FD}
%\vskip -0.3 cm
\end{figure}

The hopping parameters and the Zeeman field of the original kagome lattice are: $t_1=1$, $t_2=-0.3$, $\lambda_1=0.28$, $\lambda_2=0.2$ and $h=0.02$. We truncate the hoppings to the third nearest neighbor which are listed in Table.~\ref{tab:hopping}, where $\mu/\nu$ denote the orbitals and $i/j$ denote the sites. To fix the particle number to be half-filled, we introduce the chemical potential $\mu=0.067$. Fig.~\ref{fig:tbeff} shows a comparison between the bands of the original kagome lattice and the effective model.  \\ \newline
\begin{table}[h]
\centering
\begin{tabular}{c|c|c|c|c|c|c|c}
\hline
(m, n) &  (0, 1) & (1,0) & (1,1) & (m, n) & (0, 1) & (1,0) & (1,1) \\
\hline
$t^{00}_{m{\bf n_1}+n{\bf n_2}}$ & -0.0007+$i$0.00319 & -0.0007+$i$0.00319 & 0.0007-$i$0.00319 &$t^{11}_{m{\bf n_1}+n{\bf n_2}}$& 0.0007-$i$0.00319 & 0.0007-$i$0.00319 & -0.0007+$i$0.00319\\
\hline
(m, n) &  (2, 1) & (1,2) & (1,-1) & (m, n) &  (2, 1) & (1,2) & (1,-1) \\
\hline
$t^{00}_{m{\bf n_1}+n{\bf n_2}}$ & -0.003& -0.003 & -0.003 & $t^{11}_{m{\bf n_1}+n{\bf n_2}}$ & -0.003& -0.003 & -0.003\\
\hline
(m, n) &  (0, 2) & (2,0) & (2, 2) & (m, n) &  (0, 2) & (2,0) & (2, 2) \\
\hline
$t^{00}_{m{\bf n_1}+n{\bf n_2}}$ &-0.0019-$i$0.00046& -0.0019-$i$0.00046 & 0.0019+$i$0.00046 & $t^{11}_{m{\bf n_1}+n{\bf n_2}}$ &0.0019+$i$0.00046& 0.0019+$i$0.00046 & -0.0019-$i$0.00046\\
\hline
(m, n) &  (0, 0) & (0,1) & (-1, 0) & (m, n) &  (0, 0) & (1,0) & (0, -1) \\
\hline
$t^{01}_{m{\bf n_1}+n{\bf n_2}}$ & 0.01 & 0.01 & 0.01 & $t^{10}_{m{\bf n_1}+n{\bf n_2}}$ & 0.01 & 0.01 & 0.01 \\
\hline
(m, n) &  (1, -1) & (1,1) & (-1, -1) & (m, n) &  (-1, 1) & (-1,-1) & (1, 1) \\
\hline
$t^{01}_{m{\bf n_1}+n{\bf n_2}}$ &-0.0038 & -0.0038 & -0.0038 & $t^{10}_{m{\bf n_1}+n{\bf n_2}}$ & -0.0038 & -0.0038 & -0.0038 \\
\hline
\end{tabular}
\caption{Table of hopping parameters of the effective Hamiltonian Eq.~\ref {eq:hamiltonian_eff} appearing in the main text. }
 \label{tab:hopping}
\end{table}

\textit{Ginzburg-Landau (GL) free energy:} The GL free energy to fourth order in perturbation is given by 
\beq \nonumber
\label{FullFreeEnergy}
F[\Delta_{\bs Q_i}] &=& \alpha  \sum_i |\Delta_{\bs Q_i}|^2 + \beta_1  \sum_i |\Delta_{\bs Q_i}|^4  \\ \nonumber
&+& \beta_2 \sum_{i<j} |\Delta_{\bs Q_i}|^2|\Delta_{\bs Q_j}|^2 -  h' \gamma \sum_i\Delta_{\bs Q_i}\Delta_{-\bs Q_i} \\ \nonumber 
%&+& ...\\ \nonumber
&+& \alpha'  \sum_i |\Delta'_{\bs Q_i}|^2 + \beta_1'  \sum_i |\Delta'_{\bs Q_i}|^4  \\ \nonumber
&+& \beta_2' \sum_{i<j} |\Delta'_{\bs Q_i}|^2|\Delta'_{\bs Q_j}|^2 -h' \gamma \sum_i\Delta'_{\bs Q_i}\Delta'_{-\bs Q_i} \\ 
&+& ... \quad . 
\eeq
Here the parameters $\alpha, \alpha',  \beta_i, \beta_i', \gamma$ are constants that depend on details of the electronic structure. The expression for $\gamma$ in the flat band limit is given as 
\beq
\gamma &=& \sum_{k_n}\frac{6 h_0}{(h_0^2 + k_n^2)^2}.
\eeq
where $k_n$ is the Fermionic Matsubara frequency and $h_0$ is the static external field. The Matsubara sum can be evaluated to give
\beq
\gamma &\simeq& \frac{3 h_0}{2 T_{CO}^4} \left[ \frac{-1+ \tilde h_0^{-1}\sinh \tilde h_0}{\tilde h_0^2 (1+ \cosh \tilde h_0)} \right]\\
 \tilde h_0 &\equiv& h_0/T_{CO}.
\eeq
This expression for $\gamma$ is used in the main text of the Hamiltonian. 
 \end{document}